\documentclass[aps,pre,amsmath,amsfonts,amssymb,superscriptaddress,preprint]{revtex4}

\usepackage{amsmath}
\usepackage{amsfonts}
\usepackage{amssymb}
\usepackage[english]{babel}
\usepackage[pdftex]{graphicx}
\usepackage{tikz}
\usepackage{rotating}
\usepackage[pdftex,hypertexnames=true,linktocpage=true,breaklinks=true]{hyperref} 
\usepackage[hyphenbreaks]{breakurl}
\hypersetup{colorlinks=true,linkcolor=blue,anchorcolor=blue,citecolor=magenta,filecolor=blue,urlcolor=blue,bookmarksnumbered=false,pdfview=FitB}

\usepackage{amsthm}
\usepackage{framed}
\usepackage{amssymb}
\usepackage{amsmath,mathrsfs}
\usepackage{hhline}
\usepackage{ulem}
\usepackage{fancyhdr}
\usepackage{simplewick}

\newcommand{\be}{\begin{equation}}
\newcommand{\ee}{\end{equation}}
\newcommand{\bq}{\begin{eqnarray}}
\newcommand{\eq}{\end{eqnarray}}

\newcommand{\bv}{\bar{v}}
\newcommand{\D}{\mathrm{d}}
\newcommand{\LevelSetfunc}[2]{\Sigma_{#1}^{#2}}
\newcommand{\Mfunc}[2]{M_{#1}^{#2}}
\everymath{\displaystyle}

\begin{document}

\title{Topological origin of phase transitions in the absence of critical points of the energy landscape}

\date{\today}
\author{Matteo Gori}
\email{gori6matteo@gmail.com}
\affiliation{Aix-Marseille University, CNRS Centre de Physique Th\'eorique UMR 7332,
Campus de Luminy, Case 907, 13288 Marseille Cedex 09, France}

\author{Roberto Franzosi}
\email{roberto.franzosi@ino.it}
\affiliation{Qstar, Istituto Nazionale di Ottica, largo E. Fermi 6, 50125 Firenze, Italy}

\author{Marco Pettini}
\email{pettini@cpt.univ-mrs.fr}
\affiliation{Aix-Marseille University, CNRS Centre de Physique Th\'eorique UMR 7332,
Campus de Luminy, Case 907, 13288 Marseille Cedex 09, France}

\begin{abstract}
Different arguments led to surmise that the deep origin of phase transitions has to be identified with suitable topological changes of potential-related submanifolds of configuration space of a physical system. 
An important step forward for this approach was achieved with two theorems stating that,
for a wide class of physical systems, phase transitions should necessarily stem from topological changes of  equipotential energy submanifolds of configuration space. However, it has been recently shown that the $2D$ lattice $\phi^4$-model  provides a counterexample that falsifies the mentioned theorems. On the basis of a numerical investigation, the present work indicates the way to overcome this difficulty: in spite of the absence of critical points of the potential in correspondence of the transition energy, also the phase transition of this model stems from a change of topology of both the energy and potential level sets. But in this case the topology changes are asymptotic ($N\to\infty$). This fact is not obvious since the ${\mathbb Z}_2$ symmetry-breaking transition could be given measure-based explanations in presence of trivial topology. 

\end{abstract}
\pacs{05.20.Gg, 02.40.Vh, 05.20.- y, 05.70.- a}
\keywords{Statistical Mechanics}
\maketitle
\section{Introduction } 
For several years now, it has been put forward the idea that the relevant information about the appearance
of a phase transition in a physical system is encoded in the level sets $V_{N}(q_1,\dots,q_N) = v \in{\mathbb R}$
of its potential function $V(q_1,\dots,q_N)$, which are equivalently denoted by $\Sigma_v^{V_N}=V^{-1}_{N}(v)$ \cite{PettiniBook}. More precisely, it has been hypothesised that a phase transition has to be associated with a change of topology, at some critical value $v_c$, of these hypersurfaces of configuration space, as well as of the manifolds $\{M_{v}^{V_N}=V_N^{-1}((-\infty,v])\}_{v \in{\Bbb R}}$ bounded by the $\Sigma_v^{V_N}$.
This means that the members of the family $\{ {\Sigma_v^{V_N}\}}_{v<v_c}$ are not diffeomorphic to those of the family $\{ {\Sigma_v^{V_N}\}}_{v>v_c}$. As well, the members of the family
$\{M_{v}^{V_N}\}_{v < v_c }$ are not diffeomorphic to those of $\{ M_{v}^{V_N}\}_{v > v_c }$. 
This hypothesis came after a long conceptual pathway which has combined two aspects of Hamiltonian dynamics: 
its geometrization in terms of geodesic flows on suitably defined Riemannian manifolds, and the investigation of the dynamical counterpart of phase transitions. In fact, peculiar geometrical changes were observed to correspond to peculiar dynamical changes at a phase transition point. Then it turned out that these peculiar geometrical changes were the effect of deeper topological changes of the configuration space submanifolds $\Sigma_v^{V_{N}}$ and  $M_v^{V_N}$ \cite{CCCP,CSCFP,CCCPPG,physrep}. Then this was rigorously ascertained  for a few   exactly solvable models \cite{PettiniBook}. Finally, it was found that - for a large class of physical potentials - a phase transition \textit{necessarily} stems from the loss of diffeomorphicity of the $M_v^{V_N}$, and,  equivalently, of the $\Sigma_v^{V_N}$ \cite{prl1,TH1,TH2}. 
 This point was addressed
in Refs.\cite{prl1,TH1,TH2} where it was claimed that the occurrence of phase transitions necessarily
stems from the topological part of thermodynamic entropy.

More precisely, it has been argued that diffeomorphicity among the members of the family
$\{M_v^{V_N}\}_{v\in{\mathbb R}}$, for any $N$ larger than some $N_0$,
implies the absence of phase transitions.  

The topological approach to phase transitions has been undertaken to study a variety of systems ranging from those
undergoing entropy driven transitions \cite{carlsson1,barish} (having even applications to robotics),
to quantum phase transitions \cite{brody,BFS,volovik}, glasses and supercooled liquids
\cite{angelani,stillinger}, classical models in statistical mechanics \cite{risau,schilling,fernando}, DNA denaturation \cite{grinza}, peptide structure \cite{becker}, to quote just a few of them. 
Moreover, this unconventional (topological) viewpoint on phase transitions is of prospective interest also to tackle transitions in: \textit{i)} finite/small $N$ systems (far from the thermodynamic limit), like Bose-Einstein condensation, Dicke's superradiance in microlasers, superconductive transitions in small metallic objects;  \textit{ii)} microcanonical ensemble, especially when this is not equivalent to the canonical ensemble, and when first-order transitions are concerned;  \textit{iii)} systems without an order parameter (for example in gauge models, i.e. with local symmetries).
{\color{black}In what follows we address the $N\to\infty$ limit to join the standard definition of phase transitions, however let us stress that the study of transitional phenomena in finite $N$ systems, mentioned in the above point \textit{(i)}, is particularly relevant in many other contemporary problems, for instance related with polymers thermodynamics and biophysics \cite{bachmann}.}

Additionally, further studies in this topological framework can take advantage also of recently developed  powerful computational methods in algebraic topology, like \textit{persistent homology} \cite{Carlsson},  as shown in Ref. \cite{noi}.

{\it Remark 1}. It is worth noting that an explicit link between thermodynamics and topology is provided by the following exact formula 
\begin{equation}
S_N(v) =({k_B}/{N}) \log \left[ \int_{M_v^{V_N}}\ d^Nq\right] \nonumber
\end{equation}
\begin{equation}\label{exactS}
=\frac{k_B}{N} \log \left[ vol
[{M_v^{V_N} \setminus\bigcup_{i=1}^{{\cal N}(v)} \Gamma(x^{(i)}_c)}]\ +
\sum_{i=0}^N w_i\ \mu_i (M_v^{V_N})+ {\cal R} \right]  ,
\end{equation}
where $S_N$ is the configurational entropy, $v$ is the potential energy per degree of freedom,  and the
$\mu_i(M_v^{V_N})$ are the Morse indexes (in one-to-one correspondence with topology changes) of the
submanifolds $\{M_{v}^{V_N}=V_N^{-1}((-\infty,v])\}_{v \in{\Bbb R}}$ of configuration
space; in square brackets. The first term of Eq. \eqref{exactS} is the result of the excision of certain
neighborhoods of the critical points of the interaction potential from  $M_{v}^{V_N}$; the second term
is a weighed sum of the Morse indexes, and the third term is a smooth function of $N$ and $v$.
It is evident that sharp changes in the potential energy pattern of at least some of
the $\mu_i(M_v^{V_N})$ (thus of the way topology changes with $v$) affect $S_N(v)$ and its
derivatives. Finding adequate topology changes entailing a phase transition would provide \textit{sufficiency conditions}.

However, thus coming to the motivation of the present work, a difficulty of the theory has been recently evidenced as is discussed in the following subsection. 

\subsection{Position of the problem}

It has been recently argued \cite{kastner} against the topological theory of phase transitions on the basis of the observation that the second order phase transition of the $2D$ lattice $\phi^4$-model occurs at a critical value $v_c$ of the potential energy density which belongs to a broad interval of $v$-values void of critical points of the potential function. In other words, for any finite $N$ the $\{ \Sigma_{v<v_c}^{V_N}\}_{v \in{\Bbb R}}$ are diffeomorphic to the  $\{ \Sigma_{v>v_c}^{V_N}\}_{v \in{\Bbb R}}$ so that no topological change seems to correspond to the phase transition. This is a counterexample to the theorem in Refs. \cite{prl1,TH1}. 
A counterexample to a theorem of course falsifies it, but this can be due to two different reasons: either the proof of the given theorem is plagued by a gross mistake, or the proof is flawed in a subtler way, that is, certain logical steps have a domain of validity more restricted than implicitly assumed.  In the former case the theory is dead, in the latter case the hypotheses of the theorem need some adjustment. An instance, which is not out of place in the present context, is the famous counterexample that J. Milnor gave against De Rham's cohomology theory (the two manifolds $M = \mathbb{S}^2\times\mathbb{S}^4$, product of spheres, and $N = \mathbb{C}P^3$, complex-projective space, are neither diffeomorphic nor homeomorphic yet have the same cohomology groups, thus De Rham's theory would be inadequate to characterise the topology of manifolds). However, De Rham's cohomology theory was not discarded, and the introduction of the so called ``cup product" fixed the problem and saved the theory making it more powerful. 

Here we are in a somewhat similar situation because, as in the De Rham's case, we are in presence of the counterexample to the theorem in \cite{prl1,TH1}, given in Ref. \cite{kastner}, but nothing in Ref. \cite{kastner} has a direct relationship with the proof of the theorem, that is, it does not report on technical or logical flaws of the proof. Therefore, in order to understand whether, in analogy with the De Rham's case, our problem can be fixed by refining the basic assumptions in \cite{prl1,TH1}, we have to clarify whether or not the ${\mathbb Z}_2$ symmetry-breaking transition undergone by the 2D  lattice $\phi^4$-model corresponds to a topology change in configuration space, in spite of the absence of critical points of the potential. Thus, by addressing this point, the aim of the present paper is to understand whether the difficulty introduced by the $\phi^4$-model can be overcome. In Section \ref{tophyp} we discuss that even in the absence of critical points a sequence of diffeomorphic manifolds can have a non-diffeomorphic limit. In Section \ref{dynH} we report a remarkable outcome: on the basis of numerical simulations, and in the light of Kac's recurrence theorem, we have found evidence of an asymptotic breaking of the topological transitivity of the constant energy surfaces of the $\phi^4$-model. This splitting into two disjoint submanifolds is a major topology change which corresponds to the phase transition. This is a non-trivial result because, according to the Dobrushin-Lanford-Ruelle theory \cite{DLR,georgii},   the origin of the ${\mathbb Z}_2$ symmetry-breaking phase transition could be attributed to a loss of unicity of the statistical measure in phase space in presence of a topologically trivial support, or to a measure concentration phenomenon as suggested in \cite{kastner}. 
In Section \ref{conclus} we show that the concept of ``asymptotic diffeomorphicity'' can be given a proper mathematical definition so that, together with the results of the preceding Sections, we have found what is missing in the hypotheses of the theorems under scrutiny, hence clarifying what remains to be done to rigorously fix the problem. In Appendix the relation between the topology of the energy level sets and the potential level sets is established. 

\section{Towards a refining of the Topological Hypothesis}
\label{tophyp}

\begin{figure}[h]
\includegraphics[scale=0.55,keepaspectratio=true]{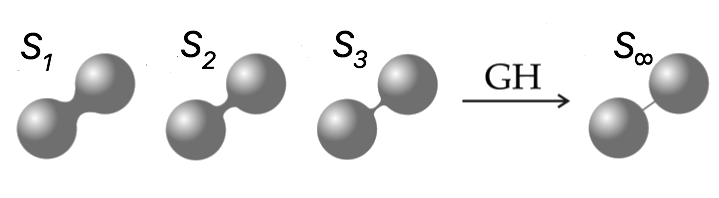}
 \caption{Gromov-Hausdorff limit of a sequence of diffeomorphic manifolds the limit of which is not diffeomorphic to the  members of the sequence.}
\label{GHlimt}
\end{figure}

Let us begin by noting that the theorem in \cite{prl1,TH1} derives uniform convergence in ${\cal C}^2([\beta(v_1),\beta(v_2)]\subset{\mathbb R})$  of the configurational Helmholtz free energy (thus the absence of first or second order phase transitions) under the assumption of diffeomorphicity at any arbitrary \textit{finite} $N\in\mathbb{N}$ of any pair of $\Sigma_{v}^N$ with $v\in [v_1,v_2]$.
While this does necessarily imply the absence of critical points of $V$ in the same interval $v\in [v_1,v_2]$, a loss of diffeomorphicity (and thus possibly  the occurrence of a phase transition) does not necessarily require the presence of critical points of the potential.
In fact, while diffeomorphicity of the level sets belonging to a so called "non-critical neck" implies the absence of critical points of any Morse function, the converse is not necessarily true. Consider, for example, a sequence of $N$-dimensional manifolds $\{{\cal{M}}_{n}\}_{n\in\mathbb{N}}$, diffeomorphic to a sphere $\mathbb{S}^N$, defined by $x_1^n+x_2^n+\dots +x_N^n=C_n$. Then as $n$ grows without bound the $\{{\cal{M}}_{n}\}$ get closer and closer to an asymptotic hypercube ${\cal{M}}_{\infty}$  no longer diffeomorphic but just homeomorphic to the ${\cal{M}}_n$, and this happens in the absence of critical points (of the functions defining the manifolds) and in presence of an unbound increase with $n$ of the mean curvature of the rounded edges of the ${\cal{M}}_n$. Another example: consider now a sequence of manifolds $\{{\cal{S}}_{n}\}_{n\in\mathbb{N}}$ obtained by successive squeezings of a sphere at any of its maximal circles thus forming a dumbbell with an increasingly tight neck as $n$ grows, hence ${\cal{S}}_{\infty}$ is homotopic to two spheres joined by an infinitely tiny cylinder as is pictorially represented in Figure \ref{GHlimt} where the so-called Gromov-Hausdorff convergence for sequences of metric spaces \cite{sormani} is pictorially displayed. Again, diffeomorphicity is asymptotically lost in the presence of an increasing mean curvature of the neck and in the absence of critical points whatsoever. 

Now, for what concerns the $2D$ lattice $\phi^4$-model, since its phase transition corresponds to an asymptotic ($N\rightarrow\infty$) spontaneous breaking of the $\mathbb{Z}_2$ symmetry, we need to clarify how this entails an asymptotic breaking of ergodicity  and hence an asymptotic change of the de Rham's cohomology group 
$H^0(\Sigma_E;{\mathbb R})$.

\section{Dynamics and topological transitivity breaking}
\label{dynH}
Here  we proceed to numerically investigate on the previously mentioned point. 
\subsection{\bf The lattice $\phi^4$ model.} The model of interest, considered in Refs.\cite{kastner,CCP},  is defined by the Hamiltonian 

\begin{equation}
{\cal H}_{N} ( p, q )= \sum_{\bf i}   \frac{p_{\bf i}^2}{2}  + V_{N}(q)
\label{Hphi_2}
\end{equation} 
where the potential function $V(q)$ is
\begin{equation}
V(q)=\sum_{{\bf i}\in{\Bbb Z}^D}\left( - \frac{\mu^2}{2} q_{\bf i}^2 +
\frac{\lambda}{4!} q_{\bf i}^4 \right) + \sum_{\langle {\bf 
ik}\rangle\in{\Bbb Z}^D}\frac{1}{2D}J (q_{\bf i}-q_{\bf k})^2\ ,
\label{potfi4}
\end{equation}
with $\langle {\bf ik}\rangle$ standing for nearest-neighbor sites on a $D$ dimensional lattice. This
system has a discrete ${\Bbb Z}_2$-symmetry and short-range
interactions; therefore, according to the Mermin--Wagner theorem,
in $D=1$ there is no phase transition whereas in $D=2$ there is a a second order 
symmetry-breaking transition,  with nonzero critical temperature, of the same
universality class of the 2D Ising model. \\
The numerical integration of the equations of motion derived from Eqs. \eqref{Hphi_2} and \eqref{potfi4} has been performed for $D=2$, with periodic boundary conditions, using a bilateral symplectic integration scheme \cite{Lapo} with time steps chosen so as to keep energy conservation within a relative precision of $\Delta E/E\simeq 10^{-6}$. The model parameters have been chosen as follows: $J=1$, $\mu^2 = 2$, and $\lambda = 3/5$. By means of standard computations as in Refs. \cite{CCCPPG} and \cite{CCP}, and for the chosen values of the parameters, the $2D$ system undergoes the symmetry-breaking phase transition at a critical energy density value  $\varepsilon_c=E_c/N\simeq 11.1$, correspondingly the critical potential energy density value is $v_c=\langle V\rangle_c/N\simeq 2.2$ \cite{nota}.  Random initial conditions have been chosen. With respect to the already performed numerical simulations we have here followed the time evolution of the order parameter (``magnetization")
\begin{equation}
M=\frac{1}{N}\sum_{\bf i} q_{\bf i}\ .
\end{equation}
This vanishes in the symmetric phase, that is for $\varepsilon > \varepsilon_c$, whereas it takes a positive or negative value in the broken symmetry phase, that is for $\varepsilon < \varepsilon_c$. However, at finite $N$ the order parameter can flip from positive to negative and viceversa. This flipping is associated with a trapping phenomenon of the phase space trajectories alternatively in one of the two subsets of the constant energy surfaces which correspond to positive and negative magnetization, respectively. This phenomenon has been investigated by computing the average trapping time $\tau_{tr}$  for different lattice sizes, and choosing values of $\varepsilon$ just below and just above $\varepsilon_c$ . 

\begin{figure}[h!]
 \centering
 \includegraphics[scale=0.45,keepaspectratio=true]{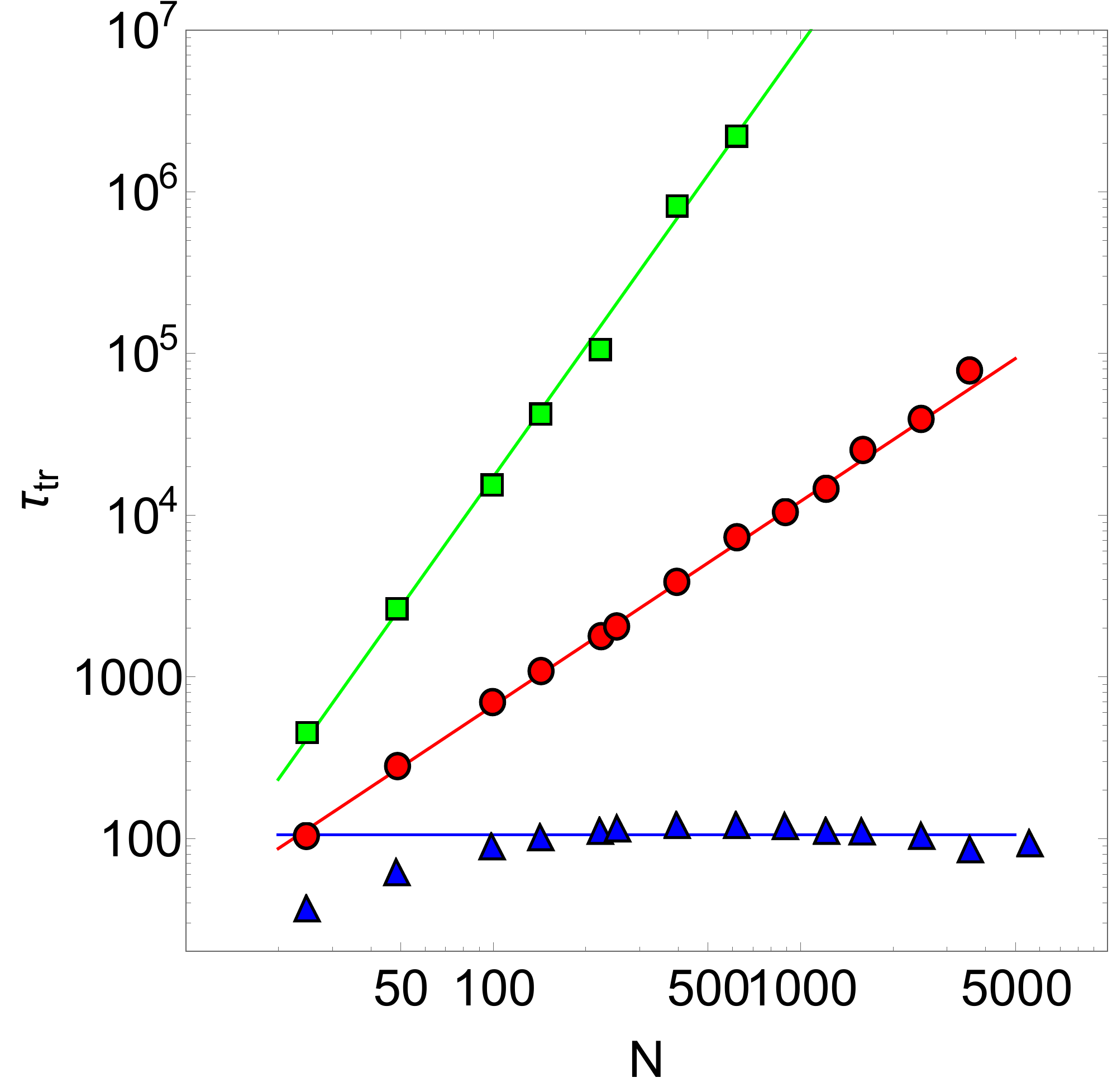}
 \caption{(Color online) Average trapping time $\tau_{tr}$ of the magnetization vs. the number of lattice sites $N$ for the 2D $\phi^4$-model. Different data series refer to different values of the energy per degree of freedom $\varepsilon$: $\varepsilon=8$ (squares), $\varepsilon=10$ (circles), both below the transition energy $\varepsilon_c=11.1$, and $\varepsilon=12$ (triangles), above the transition energy.}
\label{taueffe}
\end{figure}  

The results are displayed in Figure \ref{taueffe}. Denote with $\varphi^{{\cal H}_{N}}_t:\Sigma_E^{{\cal H}_N}\to\Sigma_E^{{\cal H}_N}$ the ${\cal H}_{N}$-flow, with $\Sigma_E^{{\cal H}_{N}}={\cal H}_{N}^{-1}(E)$ a constant energy hypersurface of phase space,  with ${\cal M}_E^+\subset\Sigma_E^{{\cal H}_{N}}$ the set of all the phase space points for which $M\ge\eta >0$,  with ${\cal M}_E^-\subset\Sigma_E$ the set of all the phase space points for which $M\le -\eta <0$, and with ${\cal M}_E^\eta\subset\Sigma_E^{{\cal H}_{N}}$ a transition region, that is, the set of all the phase space points for which $-\eta \le M\le\eta$, with $\eta\ll \langle\vert M\vert\rangle$.  Thus $\Sigma_E^{{\cal H}_{N}} ={\cal M}_E^+ \cup  {\cal M}_E^-\cup{\cal M}_E^\eta$. From the very regular functional dependences of $\tau_{tr}(N)$ reported in Figure \ref{taueffe}, we can see that:\\
\textit{At $\varepsilon < \varepsilon_c$, for any given $\tau_{tr} >0$ there exists an $N(\tau_{tr})$ such that for any $N>N(\tau_{tr})$ and $t\in [0,\tau_{tr}]$ we have   $\varphi^{{\cal H}_{N}}_t({\cal M})_E^\pm = {\cal M}_E^\pm$ }. \\ 
In other words, below the transition energy density the  subsets ${\cal M}_E^\pm$ of the constant energy surfaces 
$\Sigma_E^{{\cal H}_{N}}$ appear to be \textit{invariant} for the ${\cal H}_{N}$-flow on a finite time scale $\tau_{tr}$,
with the remarkable fact that $\tau_{tr}\to\infty$ in the limit $N\to\infty$ \cite{nota3}.
Formally this reads as 
\begin{eqnarray}
\forall A\subset{\cal M}_E^+, \forall B\subset{\cal M}_E^- &{\rm \; and\;}& t\in [0,\tau_{tr}(N)]\nonumber \\
{\rm it\, is\,} \ \   \varphi^{{\cal H}_{N}}_t(A)\cap B &=& \emptyset \ .
\label{TT1}
\end{eqnarray} 
To the contrary:\\
\textit{At $\varepsilon > \varepsilon_c$, there exists a $\tau_{tr}^0 >0$ such that for any $N$ and }
\begin{eqnarray}
\forall A\subset{\cal M}_E^+, \forall B\subset{\cal M}_E^- &{\rm \; and\;}& t>\tau_{tr}^0 \nonumber \\
{\rm it\, is\,} \ \   \varphi^{{\cal H}_{N}}_t(A)\cap B &\neq & \emptyset \ .
\label{TT2}
\end{eqnarray} 
The divergence of $\tau_{tr}$ with $N\to\infty$ - below the transition point - has two remarkable consequences. The first one is related with Kac's recurrence theorem \cite{kac}, which, applied here, means that for the ergodic transformation $\varphi^{{\cal H}_{N}}_t:\Sigma_E^{{\cal H}_{N}}\to\Sigma_E^{{\cal H}_{N}}$ \cite{notaergodicity} and its positive invariant measure $d\mu =d\sigma /\Vert\nabla {\cal H}_N\Vert$, where $d\sigma$ is the maximal form on $\Sigma_E^{{\cal H}_{N}}$, the first return map for  ${\cal M}_E^\eta$ defined as
\[
n({{\cal M}_E^\eta}) = \min \{n\ge 1 : [\varphi^{{\cal H}_N}_{\Delta t}]^n(x)\in {{\cal M}_E^\eta} ,\  \forall x\in {{\cal M}_E^\eta}\}
\]
can be integrated. And, having normalized to 1 the measure of $\Sigma_E^{{\cal H}_{N}}$, the mean return time to ${\cal M}_E^\eta$ is \cite{varandas}
\begin{equation}\label{mu-diverg}
\int_{{\cal M}_E^\eta} n(x\, ; {{\cal M}_E^\eta})\frac{ d\sigma} {\Vert\nabla {\cal H}_{N}\Vert} = \frac{1}{\mu ({\cal M}_E^\eta )}
\end{equation} 
and since the return time in ${\cal M}_E^\eta$ is given by the flipping time between opposite values of the magnetisation, the return time coincides with $\tau_{tr}$, thus the divergence of $\tau_{tr}$ entails the vanishing of the measure of the transition region ${\cal M}_E^\eta$.

{\it Remark 2}. Kac's theorem concerns a discrete dynamics, which is actually the case of the numerical simulation of a Hamiltonian flow; however, the numerical algorithm is a symplectic mapping producing pseudo orbits in phase space which, after Moser's interpolation theorem \cite{moser,benettin}, are homeomorphic to true phase space trajectories via an homeomorphism which can be arbitrarily close to the identity according to the value of $\Delta t$.

{\it Remark 3}.  Even though $N=5000$ is far from the limit $N\to\infty$, (which is also true for any arbitrarily large but finite number), we can reasonably expect that the power-law-divergences of $\tau_{tr}$ are stable with $N$ because any increase of the number of lattice sites  $N$ can be thought of as the result of  glueing together an arbitrary number of replicas of a given smaller lattice, by keeping constant the energy density. In fact, consider for example the point (red circle) in \autoref{taueffe} which corresponds to $N=50$, then by glueing together four such lattices we obtain a lattice of $N=200$ sites. These two points define a line of a given slope (power law). Then consider the point at $N=800$, which can be thought of as the result of glueing together four lattices of $N=200$, this point belongs to the line defined by the preceding two. Then consider the point at $N=3200$, which can be thought of as the result of glueing together four lattices of $N=800$, also this point belongs to the same line; in other words, we see that the power law of $\tau_{tr}(N)$ does not change by increasing $N$. Since there is no reason for interrupting such a progression, we can reasonably  assume that the observed power laws of $\tau_{tr}(N)$ hold true at any $N$.

The second, and related, consequence of the divergence of $\tau_{tr}$ with $N\to\infty$ is that  \textit{topological transitivity}  \cite{toptran} of $\Sigma_E^{{\cal H}_{N}}$ is broken on the timescale 
${\tau}_{tr}$. 
On the contrary, at $\varepsilon > \varepsilon_c$, the trapping time is short (in units of the inverse of the shortest linear frequency of the system) and is a flat function of $N$, so that, after a timescale $\tau_{tr}^0 >0$ independent of $N$, the dynamically evolved set $\varphi^{{\cal H}_{N}}_{t>{\tau}_{tr}^0}(A)$ of $A$ always overlaps with $B$:
above the transition energy the $\Sigma_E^{{\cal H}_{N}}$ are \textit{topologically transitive}.
 
The divergence of $\tau_{tr}(N)$ in the limit $N\to\infty$ - at $\varepsilon < \varepsilon_c$ - is thus equivalent to the asymptotic breaking of topological transitivity. Moreover, on metric and compact topological spaces,  topological transitivity is equivalent to \textit{connectedness} of the space \cite{toptran}, so the loss of topological transitivity entails the loss of connectedness, that is, \textit{a major topological change} of the space. Let us denote by $H_{\tau}^0(\Sigma^{{\cal H}_{N}}_E;{\Bbb R})$ the ``finite time zeroth cohomology space" of any given $\Sigma_E^{{\cal H}_{N}}$.  The dimension of this cohomology space (the Betti number $b_0$) counts the number of connected components of $\Sigma_E^{{\cal H}_{N}}$ and is invariant under diffeomorphisms of the $\Sigma_E^{{\cal H}_{N}}$. At $\varepsilon < \varepsilon_c$, for any $\tau < \tau_{tr}(N)$ it is $b_0=\dim H_{\tau}^0(\Sigma^{{\cal H}_{N}}_E;{\Bbb R}) = 2$,  and, at $\varepsilon > \varepsilon_c$, $b_0=\dim H_{\tau}^0(\Sigma^{{\cal H}_{N}}_E;{\Bbb R}) = 1$.  Hence the asymptotic jump of a diffeomorphism-invariant ($b_0$) across the phase transition point, which can be deduced by our numerical computations, means that the $\Sigma^{{\cal H}_{N}}_E$ undergo an \textit{asymptotic loss of diffeomorphicity}, in the absence of critical points  of the potential $V_N(q)$. 

Since $\Sigma_E^{{\cal H}_{N}} ={\cal M}_E^+ \cup  {\cal M}_E^-\cup{\cal M}_E^\eta$, and since the residence times in the transition region are found to be very short and independent of $N$ - so that the relative measure ${\sl meas}({\cal M}_E^\eta)/{\sl meas}({\cal M}_E^\pm)$ vanishes in the limit $N\to\infty$ - Eq. \eqref{TT1} means that below the transition energy the \textit{topological transitivity} of $\Sigma_E^{{\cal H}_{N}}$ is broken up to a time $\tau_{tr}(N)$ --  which is divergent with $N$. To the contrary, above the transition energy the $\Sigma_E^{{\cal H}_{N}}$ are \textit{topologically transitive} \cite{toptran}.  The asymptotic breaking of topological transitivity at $\varepsilon < \varepsilon_c$, that is the divergence of $\tau_{tr}(N)$ in the limit $N\to\infty$,  goes together with asymptotic ergodicity breaking due to the ${\Bbb Z}_2$-symmetry breaking.

\subsection{Absence of measure concentration}
\begin{figure}[h]
 \centering
 \includegraphics[scale=0.43,height=5cm,width=9cm]{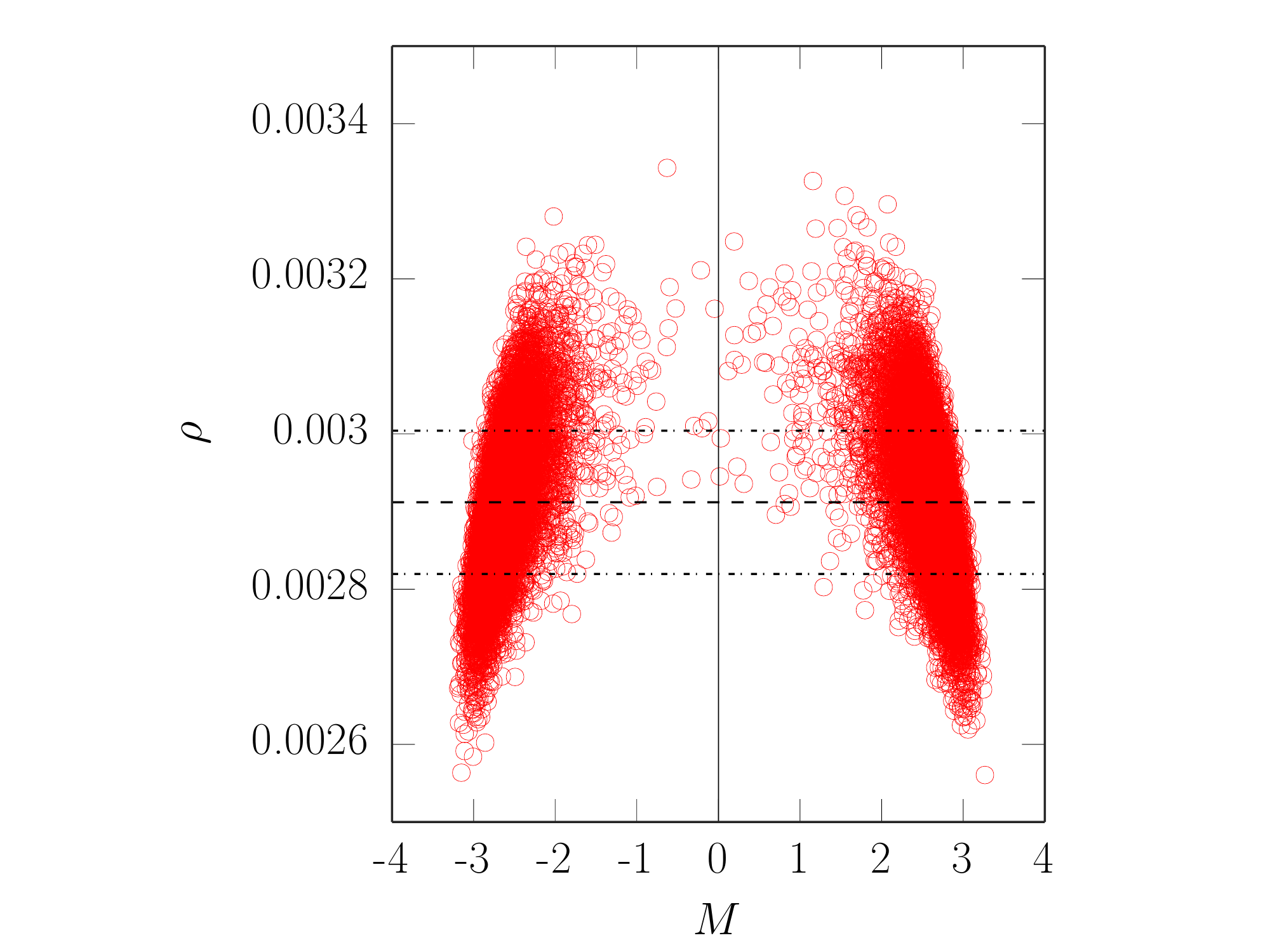}
  \includegraphics[scale=0.43,height=5cm,width=9cm]{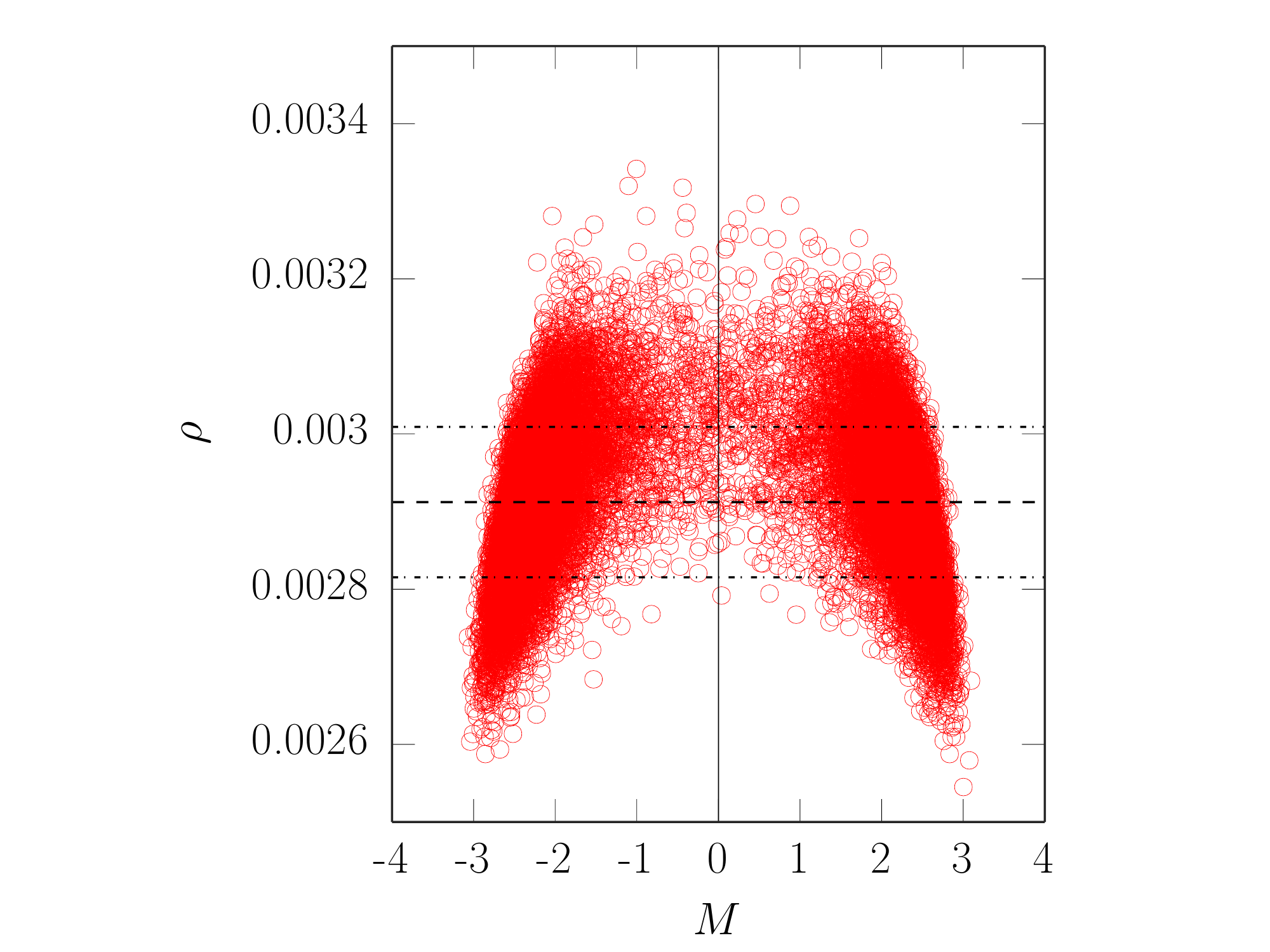}
\includegraphics[scale=0.43,height=5cm,width=9cm]{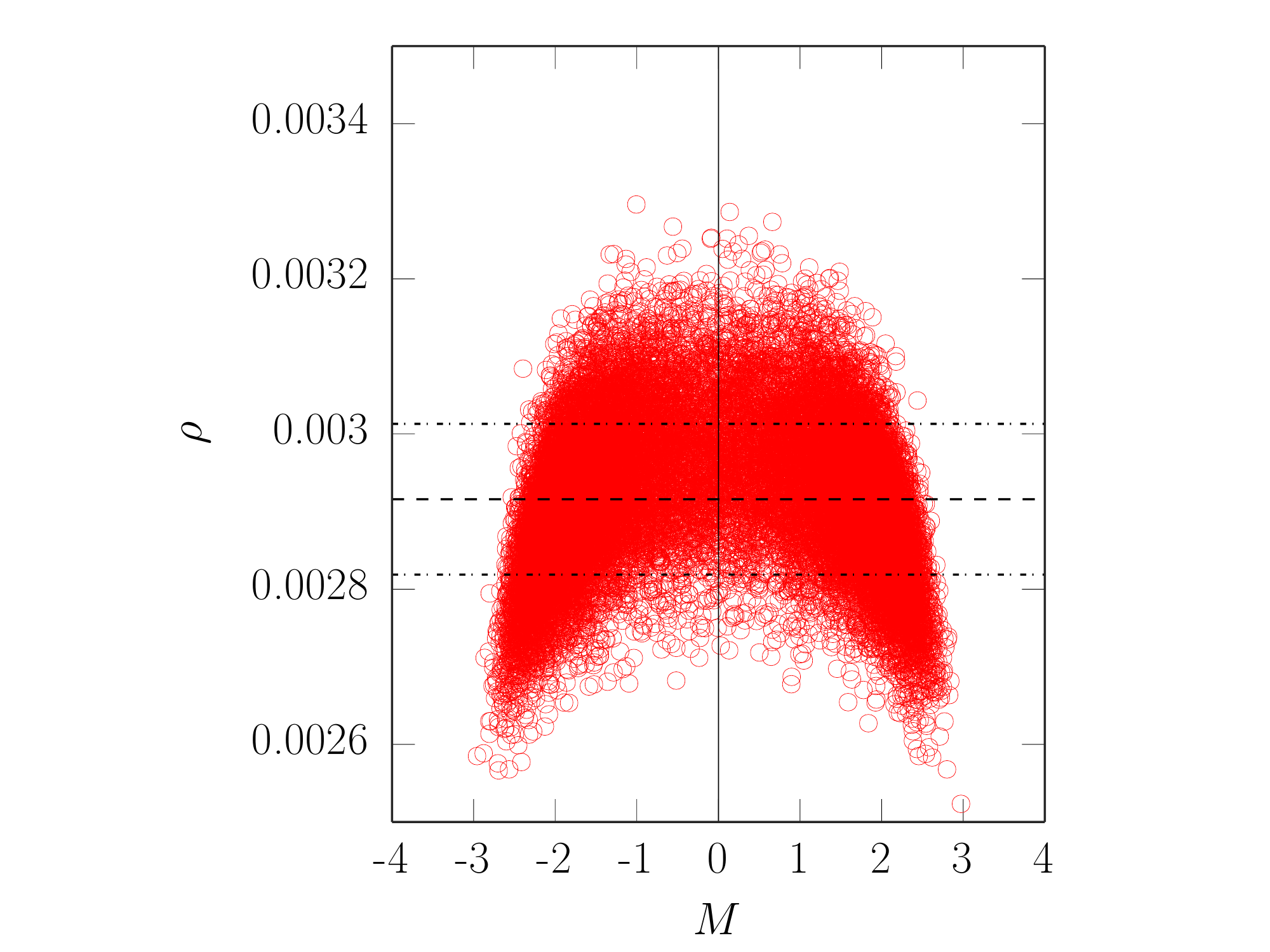}
 \caption{(Color online) Dynamical sampling of point values of $\rho =1/\Vert\nabla {\cal H}_{N}\Vert$ versus point values of the  magnetization $M$. Data refer to $32\times 32$ square lattice. The upper panel corresponds to  $\varepsilon=10$, below the phase transition; the intermediate panel corresponds to  $\varepsilon=11\simeq \varepsilon_c$, at the phase transition; lower panel corresponds to  $\varepsilon =12$, above the transition energy. The horizontal dashed lines mark the averages of $\chi$ computed on the sets of points displayed. The dot-dashed lines mark the standard deviations of the same sets of points.}
\label{noconcentr}
\end{figure}

Finally, to confirm the interpretation given to the previously reported phenomenology, we have ruled out an alternative possibility which might be considered in analogy with the Dobrushin-Lanford-Ruelle theory, already mentioned in the Introduction, that is, one has to exclude that the observed ergodicity breaking stems from a concentration phenomenon \cite{ledoux} - on a topologically trivial phase space -
of the microcanonical measure 
\begin{equation}
\mu(A)=\int_{A}\, d\mu = \int_{A}\, \dfrac{d\sigma}{\Vert\nabla {\cal H}_N\Vert }\qquad A\subset \Sigma^{{\cal H}_{N}}_{E}
\end{equation}
where $d\sigma$ is the area measure of regular level sets in euclidean space (in the following we will equivalently use ``area'' and ``volume'' referring to the integration with the respect to the differential form of maximal rank on the level sets). Saying it informally,
in principle at increasing $N$ the microcanonical measure could split and better and better
concentrate on two disjoint subsets of the corresponding $\Sigma^{{\cal H}_{N}}_E$.
And this  could be due to the "weight" function $\rho=\|\nabla \mathcal{H}_{N}\|^{-1}$
polarizing the measure also in the absence of a modification of the shapes of the energy
level sets. If this was the case then the values taken by $\chi$ should be very small when $M\simeq 0$, and comparatively very large when $M\simeq \langle M_\pm\rangle$.
In order to exclude this possibility, we have proceeded as follows. We have numerically computed the point values of $\rho$ along the dynamics and recorded them as a function of the corresponding point values of the magnetization $M$. In Figures \ref{noconcentr} we report the outcomes of numerical computations which go just in the opposite direction, in fact the $\rho$ values are mostly concentrated in
the same (small) interval almost independently of the values of $M$ both above and below the transition energy.
This means that the relative measure of the subsets ${\cal M}_E^\pm$ and  ${\cal M}_E^\eta$ is affected by the shape of the $\Sigma_{E}$ (through the euclidean area measure) instead of by the values of the density function $\rho$.\\
Notice that at  $\varepsilon=10$, below the transition energy density $\varepsilon_c=11.1$, the rarefaction of points in the interval $M\ge -2$ and $M\le 2$ is due to a short residence time in the transition region of phase space, in agreement with the remark that ${\sl meas}({\cal M}_E^\eta)/{\sl meas}({\cal M}_E^\pm)$ decreases with growing $N$, as discussed in the preceding Section. In other words, the crowding of points around two clouds of opposite values of $M$, together with the independence of $\chi$ on $M$, confirms that at increasing $N$ the phase space tends to break into two disjoint regions bridged by a thinner and thinner  region. 

\subsection{Geometric signature of the neck on the $\Sigma_E^{\mathcal{H}_N}$ of phase space}
In the absence of the phenomenon of measure concentration, and mainly after the implication of Kac's recurrence theorem,
we are led to think that, below the transition energy,  a ``neck'' exists in phase space, joining two otherwise disjoint
regions of the energy level sets. We call this transition region a ``neck'' because after Eq.\eqref{mu-diverg} its measure  vanishes asymptotically, that is, this region shrinks at increasing $N$. A high dimensional neck in a high dimensional space
is far from direct intuition, nevertheless we can try to characterise it through some geometric observable. In order to do
this we proceed as follows.
Let us note that, in the absence of critical points in an interval $[a,b]$, the explicit form of the diffeomorphism $\boldsymbol{\zeta}$ that maps one to the other the level sets $\Sigma_c^{f}=f^{-1}(c)$, $c\in[a,b]$, of a function $f:{\Bbb R}^N\to{\Bbb R}$ is explicitly given by \cite{hirsch}
\begin{equation}
\dfrac{\D x^i}{\D c} =\dfrac{\nabla^i f(x)}{\Vert\nabla f(x)\Vert^2}\,.
\label{hirschflow}
\end{equation}
This applies as well to the energy level sets $\Sigma_{E}^{\mathcal{H}_N}$ in phase space; in this case,
the vector fields that generates the diffeomorphism is
\begin{equation}
\dfrac{\D x^i}{\D E}=\zeta^{i}(x)=\dfrac{\nabla^{i} {\cal H}_N}{\|\nabla {\cal H}_N\|^2}\quad i=1,...,2N
\label{hirschflowHam}
\end{equation}
where $x^{i}=p^{i}$ and $x^{N+i}=q^{i}=\phi^{i}$.
If we consider an infinitesimal change of the energy $E\rightarrow E+\delta E$ with $\vert\delta E\vert/E\ll 1$, and denote with $\delta (x)$ the field of local distances  between two level sets $\Sigma_E$ and $\Sigma_{E+\delta E}$, from 
$x^i(E+\delta E)=x^i(E) + \zeta^i \delta E$ and using Eq.\eqref{hirschflowHam} at first order in $\delta E$,  we get $\delta(x)=\delta E/\Vert \nabla {\cal H}_{N}\Vert_x=\rho(x)\ \delta E$. 
Moreover the divergence $\mathrm{div}\boldsymbol{\zeta}$ in euclidean configuration space can be related with the variation rate of the measure of the  microcanonical area  $\mathrm{d}\mu=\rho\mathrm{d}\sigma$ over regular level sets $\Sigma_E^{\mathcal{H}_N}$.
\begin{figure}[h]
 \centering
 \includegraphics[height=9cm,width=14cm]{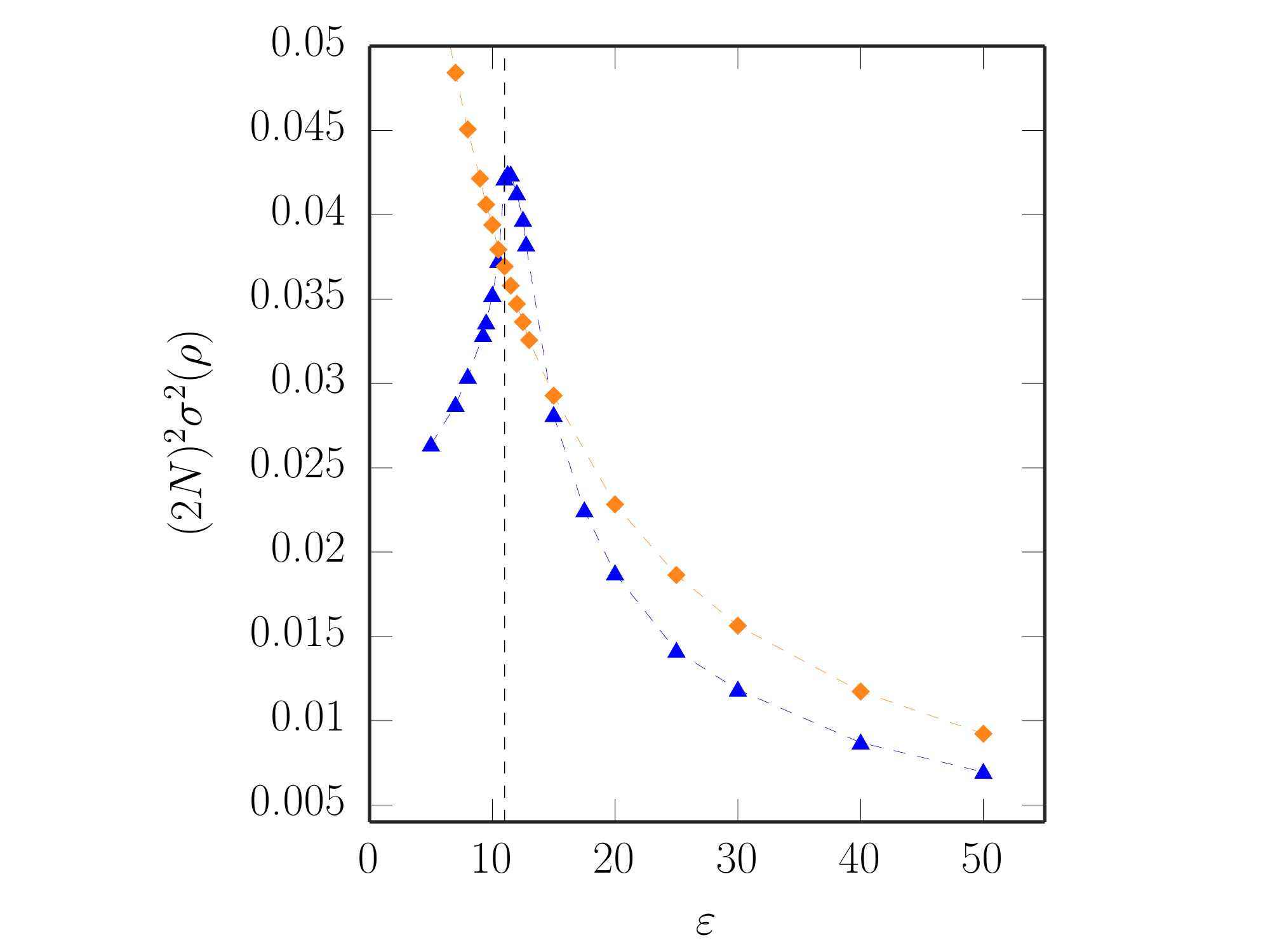}
\caption{(Color online) Variance of $\rho = 1/\Vert \nabla {\cal H}_{N}\Vert$ vs. total energy per degree of freedom $\varepsilon$ for 1D and 2D $\phi^4$-models, and for lattice sizes: $N=30\times 30$ (triangles) in the 2D case, and $N=900$ (rhombs) in the 1D case. The vertical dashed line indicates the phase transition point at $\varepsilon\simeq 11.1$.}
\label{var-chiH}
\end{figure}

\begin{figure}[h!]
 \centering
 \includegraphics[height=9cm,width=14cm]{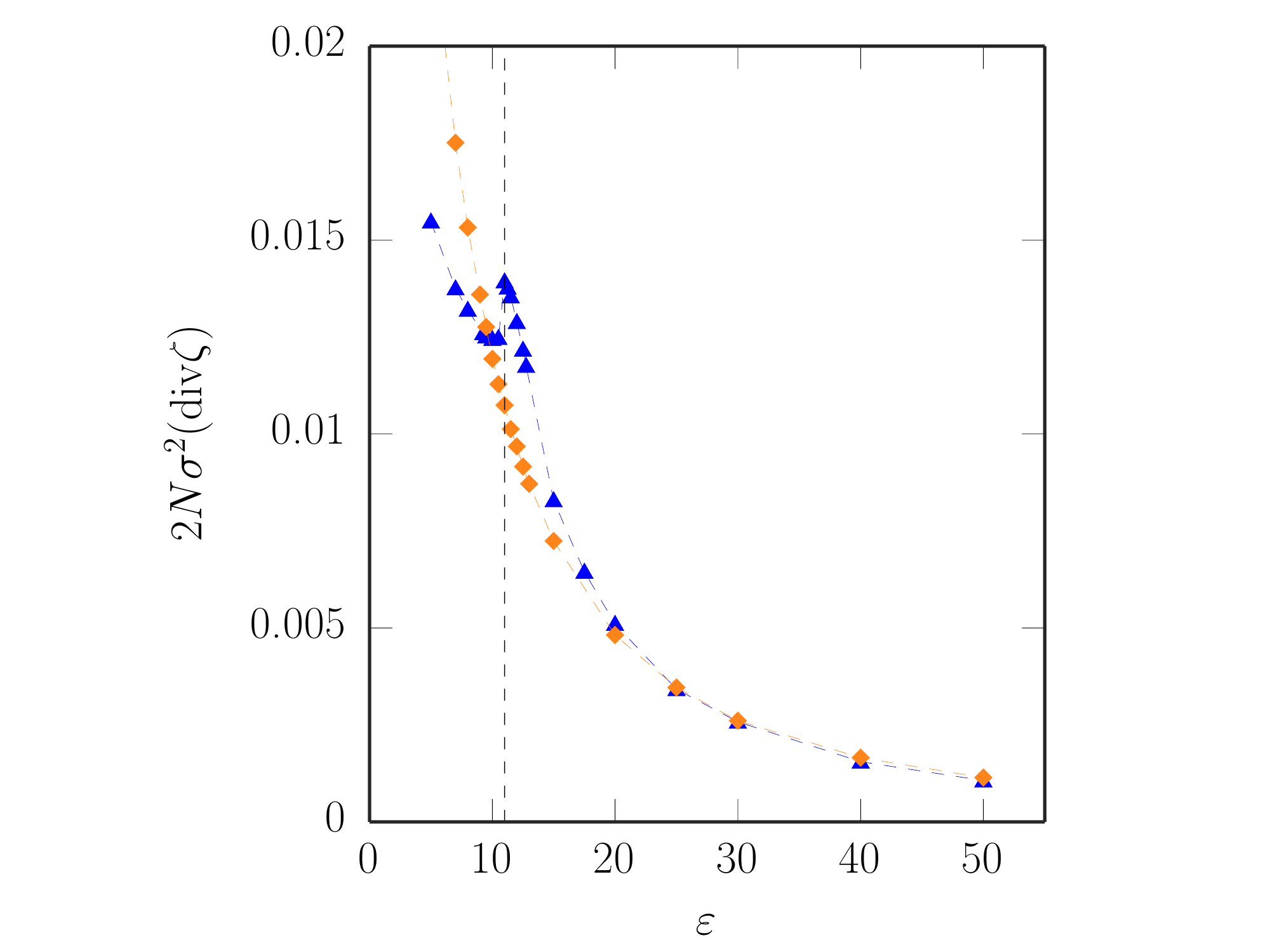}
\caption{(Color online) Variance of $\mathrm{div}\boldsymbol{\zeta}$, where $\boldsymbol{\zeta}=\nabla{\cal H}_{N}/\Vert\nabla{\cal H}_{N}\Vert^2$, vs. total energy per degree of freedom $\varepsilon$ for 1D and 2D $\phi^4$-models, and for lattice sizes: $N=30\times 30$ (triangles) in the 2D case, and $N=900$ (rhombs) in the 1D case. The vertical dashed line indicates the phase transition point at $\varepsilon\simeq 11.1$.}
\label{var-divergH}
\end{figure}

The first variation formula for the induced measure of the Riemannian area  $\mathrm{d}\sigma$ along the flow $x(E)$ reads \cite{LeeGeometry}:
\begin{equation}
\mathrm{d}\sigma(x(E+\delta E))=\left(1-\rho M_1 \delta E\right)\mathrm{d}\sigma(x(E))+o(\delta E)
\end{equation}
where $M_1$ is the sum of the principal (extrinsic) curvatures of $\Sigma_E^{\mathcal{H}_N}$ 
that is given by
\begin{equation}\label{m1E}
M_1=-\mathrm{div}\left(\dfrac{\nabla {\cal H}_{N}}{\|\nabla {\cal H}_{N}\|}\right).
\end{equation}
Applying the Leibnitz rule, the first variation formula for the measure of the microcanonical area is
\begin{equation}\label{muareaH}
\begin{split}
&\mathrm{d}\mu(x(E+\delta E))=\rho(x(E+\delta E))\mathrm{d}\sigma(E+\delta E)=\\
&=\left[1+\left(-\rho M_1+\dfrac{(\nabla^i {\cal H}_{N})}{\|\nabla {\cal H}_{N}\|}\nabla_i\rho\right) \delta E \right]\mathrm{d}\mu=\\
&=\left(1+\mathrm{div}\boldsymbol{\zeta}  \delta E\right)\mathrm{d}\mu(x(E))
\end{split}
\end{equation}
Then, the two following quantities have been numerically computed along the Hamiltonian flow:
 $\sigma^2(\rho)=\langle\rho^2\rangle_{\Sigma_E^{{\cal H}_{N}}} - \langle\rho\rangle^2_{\Sigma_E^{{\cal H}_{N}}}$ and $\sigma^2(\mathrm{div}\boldsymbol{\zeta})=\langle(\mathrm{div}\boldsymbol{\zeta})^2\rangle_{\Sigma_E^{{\cal H}_{N}}} - \langle(\mathrm{div}\boldsymbol{\zeta})\rangle^2_{\Sigma_E^{\mathcal{H}_N}}$.
These are functions of $N$ and of the specific energy ${\varepsilon}=E/N$.
The outcomes, reported in  \autoref{var-chiH} and \autoref{var-divergH}, show very different patterns in the $1D$ and $2D$ cases: monotonic  for the $1D$ case, non-monotonic for  the $2D$ case, displaying cuspy points at ${\varepsilon}={\varepsilon}_c$ (the phase transition point) of $\sigma^2(\rho)$ and of $\sigma^2(\mathrm{div}\boldsymbol{\zeta})$. As $\rho =1/\Vert\nabla {\cal H}_{N}\Vert$ is locally proportional to the distance between nearby level sets, its variance is a measure of the total dishomogeneity of this distance, so that a peak of $\sigma^2(\rho)$ can be reasonably attributed to the presence of the ``neck" in the $\{\Sigma_E^{\mathcal{H}_N} \}_{E <E_c}$ foliation of phase space. The same is true for $\sigma^2(\mathrm{div}\boldsymbol{\zeta})$ since $\mathrm{div}\boldsymbol{\zeta}$ is locally proportional to the variation of the area of a small surface element when a level set is transformed into a nearby one by the diffeomorphism in Eq. \eqref{hirschflow}.

The result displayed in \autoref{var-divergH} can be given also another geometrical meaning, that is, $\sigma^2(\mathrm{div}\boldsymbol{\zeta})$ also measures the total variance of the mean curvature of a level set (``total'' meaning integrated on the whole manifold). This follows from the possible interpretation of the divergence of the vector field $\boldsymbol{\zeta}$ as a curvature property of the level sets $\Sigma_{E}^{{\cal H}_{N}}$ of the Hamiltonian function ${\cal H}_{N}$ in phase space, or, alternatively, of the level sets $\Sigma_{v}^{V_N}$ of the potential function $V_{N}$ in configuration space.
To show this, we start by pointing out that statistical mechanics in the microcanonical ensemble is invariant for volume-preserving diffeomorphisms. In fact, in the microcanonical ensemble the thermodynamics is derived from the entropy, defined as $S_N(E)=k_B \log\Omega_N(E)$ with 

\begin{equation}
\begin{split}
&\Omega_N(E)=\dfrac{\D}{\D E}\int \Theta\left({\cal H}_{N}(\mathbf{p},\mathbf{q})- E\right) \D\mathrm{Vol}_{g}\\
&=\dfrac{\D}{\D E}\int \Theta\left(H_{N}(\mathbf{p},\mathbf{q})- E\right) \left|\mathrm{det} g\right|^{1/2}\prod_{i=1}^N\D p_i \D q_i \\
&=\int_{\Sigma_E^{{\cal H}_{N}}} \rho\,\,\D\sigma_{g}
\end{split}
\end{equation}

where $g$ is any Riemannian metric for phase space and $\D\mathrm{Vol}_{g}$ the corresponding volume element giving the same total volume of the Liouville-Lebesgue measure. Thus  equivalent thermodynamic descriptions of the same system are  
given by any two metrics $g$ and $\widetilde{g}$ such that $\left|\mathrm{det} g\right|= \left|\mathrm{det} \widetilde{g}\right|$, being both preserving the symplectic volume.
This arbitrariness can be used to chose a rescaled metric $\tilde{g}$ that encodes
in the Riemannian structure the information on the  density of states given by the function $\rho = \|\boldsymbol{\zeta}\|_{g}=\|\nabla {\cal H}_{N}\|_{g}^{-1}$. This procedure is inspired by works on  ``manifolds with density'' \cite{gromov,morgan,bayle}.
In particular for a given metric $g$ we look for a new metric $\widetilde{g}$ such that 
$\D \mathrm{Vol}_{g}=\D \mathrm{Vol}_{\widetilde{g}}$ and 
\begin{itemize}
\item $\boldsymbol{\zeta}$ is the vector field normal to the level sets 
\begin{equation}
\widetilde{g}\left(\boldsymbol{\zeta},\boldsymbol{\zeta}\right)=\|\boldsymbol{\zeta}\|_{\widetilde{g}}=1\,\,\,;
\end{equation}
\item the Riemannian volume form $\D \sigma_{\widetilde{g}}$ induced on the level
sets of the Hamiltonian function is the microcanonical volume form with metric $g$, that is
\begin{equation}
\D \sigma_{\widetilde{g}}=\chi\D \sigma_{g} = \D \mu\,.
\end{equation}
\end{itemize}
where $\D \mu$ is the measure in Eq.\eqref{muareaH}. A possible choice for such a metric $\tilde{g}$ is
\begin{equation}
\begin{cases}
& \widetilde{g}\left(\boldsymbol{N},\boldsymbol{N}\right)=\rho^{-2}g\left(\boldsymbol{N},\boldsymbol{N}\right)\\
& \widetilde{g}\left(\boldsymbol{N},\boldsymbol{X}\right)=g\left(\boldsymbol{N},\boldsymbol{X}\right)=0\\
& \widetilde{g}\left(\boldsymbol{X},\boldsymbol{Y}\right)=\rho^{2/(N-1)}g\left(\boldsymbol{X},\boldsymbol{Y}\right)\,.\\
\end{cases}
\end{equation}
where $\boldsymbol{N}$ is a vector field orthogonal to the level sets of the  Hamiltonian and 
$\boldsymbol{X},\boldsymbol{Y}$ are vector fields tangent to the same level sets.
With this metric the microcanonical volume can be written as
\begin{equation}
\Omega(E)=\int_{\Sigma_E^{{\cal H}_{N}} }\,\D\sigma_{\widetilde{g}}
\end{equation}
which is just the measure of the geometric volume of the $\Sigma_{E^{\cal H}_{N}}$ (or, equivalently, the area of the hypersurface). Let us recall that the sum of the principal curvatures  $\widetilde{M}_1$ (that is $N$ times the mean curvature), for the $\Sigma_E^{{\cal H}_{N}}$, 
embedded in the phase space endowed with the metric $\widetilde{g}$, is related with the Lie derivative of the Riemannian area form on regular level sets with respect to normal vector field $\widetilde{\boldsymbol{\nu}}$ by
\begin{equation}
\widetilde{M}_{1}\,\,\D\sigma_{\widetilde{g}}=-\mathcal{L}_{\widetilde{\boldsymbol{\nu}}}\D\sigma_{\widetilde{g}}=-\mathcal{L}_{\boldsymbol{\zeta}}\D\sigma_{\widetilde{g}}.
\end{equation}
According to Eq.\eqref{muareaH}, the Lie derivative of the volume form $\D\sigma_{\widetilde{g}}$ can be expressed as a function
of $\mathrm{div}_{g} \boldsymbol{\zeta}$
\begin{equation}
\mathcal{L}_{\boldsymbol{\zeta}}\D\sigma_{\widetilde{g}}=\mathcal{L}_{\boldsymbol{\zeta}}\D \mu=\mathrm{div}_{g}\boldsymbol{\zeta}\D \mu\,\,.
\end{equation}
It follows that the divergence of the vector field $\boldsymbol{\zeta}$ in the phase space endowed with the metric $g$ can be interpreted as the opposite of  the sum of principal curvatures  $\widetilde{M}_{1}$ of the energy level sets $\Sigma_E^{{\cal H}_{N}}$ embedded in the phase space endowed with the metric $\widetilde{g}$
\begin{equation}
\mathrm{div}_{g}\boldsymbol{\zeta}=-\widetilde{M}_1
\end{equation}
In conclusion, because of an apparent discontinuity - and of course within the limits proper to a numerical result -  the pattern of $\sigma^2(\mathrm{div}\boldsymbol{\zeta})$ versus energy can be defined ``singular'', thus also the way of changing of the variance of the total curvature (thus the way of changing of geometry) of the energy level sets at the phase transition point can be defined ``singular''.
 
To help intuition, consider the limiting case of a sequence of isotropic manifolds, say spheres foliating an embedding space, then imagine that - by changing a parameter that labels the leaves of the foliation - a neck (like in the dumbbells in \autoref{GHlimt}) suddenly appears. The variance of the total sum of principal curvatures of course vanishes for the spheres whereas the necks bring about regions of negative principal curvatures entailing local variations of the mean curvature and, consequently, the sudden appearance of a non-vanishing variance of the total mean curvature. Whence a discontinuous pattern of the total variance of the mean curvature can be intuitively associated with an abrupt geometrical change of the leaves of the foliation.

\subsection{Geometric signature of the neck of the $\Sigma_v^{V_N}$ in configuration space}
The breaking of topological transitivity of the $\{\Sigma_E^{{\cal H}_{N}}\}_{N\in\mathbb{N}}$ implies the same phenomenon for configuration space and its potential level sets submanifolds $\{\Sigma_v^{V_N}=V^{-1}_N(v)\}_{N\in\mathbb{N}}$ (see Appendix). These level sets are the basic objects, foliating configuration space, that enter the theorems in \cite{prl1,TH1,TH2}, and represent the topologically nontrivial part of phase space. The link of these geometric objects with microcanonical entropy is given by 
\begin{equation}\label{mu-ent}
\begin{split}
S(E) =&\frac{k_B}{2N} \log \int_0^E d\eta \int d^Np\  \delta \left(\sum_{\bf i} p_{\bf i}^2/2 - \eta\right)\times\\
&\times\int_{\Sigma_{E-\eta}^{V_{N}}} \frac{d\sigma}{\Vert\nabla V_{N}\Vert} \,\,.
\end{split}
\end{equation}
As $N$ increases the microscopic configurations giving a relevant contribution to the entropy, and to any  microcanonical average, concentrate closer and closer on the level set $\Sigma_{\langle E-\eta\rangle}^{V_N}$. Therefore, it is interesting to make a direct numerical analysis on these level sets at different $N$ values to find out - with a purely geometric glance - how configuration space asymptotically breaks into two disjoint components. The intuitive picture is that, approaching from above ($\varepsilon > \varepsilon_c$) the transition point,  some subset of each $\Sigma_v^{V_N}$, a ``high dimensional neck", should be formed also in configuration space  bridging the two regions ${\cal M}_v^+$ and ${\cal M}_v^-$. And this neck should increasingly shrink with increasing $N$. To perform this analysis we resort to a Monte Carlo algorithm constrained on any given $\Sigma_v^{V_N}$. This is obtained by  generating a Markov Chain with a Metropolis importance sampling of the weight $\chi=1/{\Vert\nabla V_{N}\Vert}$. Then we proceed by computing the same geometric quantities that have been computed in phase space. 

We can now repeat almost verbatim for configuration space what has been discussed above for phase space.
Consider an infinitesimal change of potential energy $v\rightarrow v+\epsilon_{v}$ with $\vert\epsilon_{v}\vert/v \ll 1$, and denote with $\delta (q)$ the field of local distances  between two level sets $\Sigma_v$ and $\Sigma_{v+\epsilon_{v}}$, from 
$q^i(v+\epsilon_{v})=q^i(v) + \xi^i \epsilon_{v}$ and using Eq.\eqref{hirschflow}, at first order in $\epsilon_{v}$,  we get $\delta(q)=\epsilon_{v}/\Vert \nabla V_{N}\Vert_q=\epsilon_{v}\chi(q)$. 
Again the divergence $\mathrm{div}\boldsymbol{\xi}$ in euclidean configuration space can be related with the variation rate of the measure of the  configurational microcanonical area  $\mathrm{d}\mu=\chi\mathrm{d}\sigma$ over regular level sets $\Sigma_{v}^{V_N}$, now with $\boldsymbol{\xi}=\nabla V_{N}/\Vert\nabla V_{N}\Vert^2$ and $\chi = 1/\Vert\nabla V_{N}\Vert$.

\begin{figure}[h]
 \centering
 \includegraphics[scale=0.45,keepaspectratio=true]{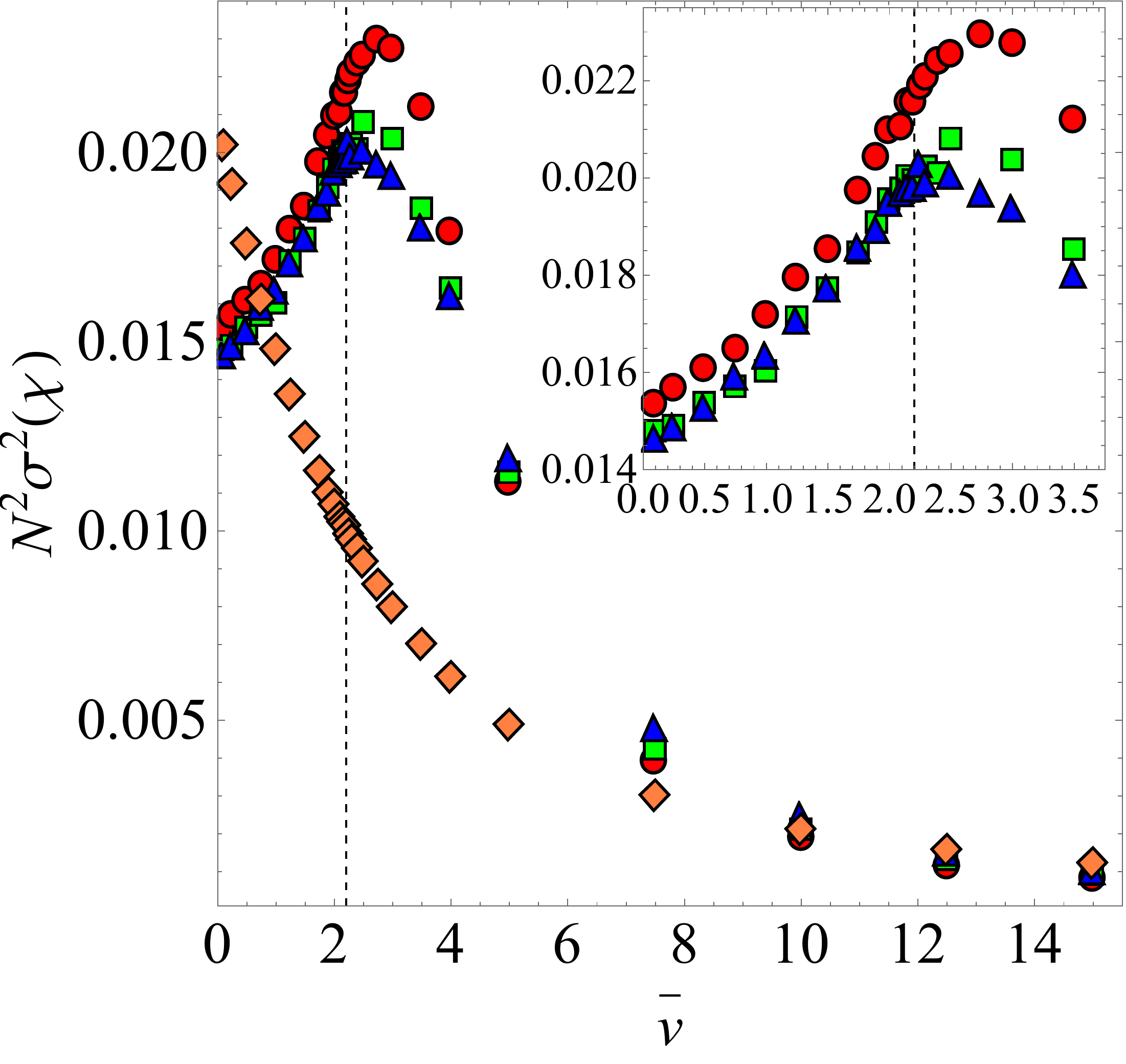}
\caption{(Color online) Variance of $\chi =1/\Vert\nabla V_{N}\Vert$ vs.  potential energy per degree of freedom $\bar{v}$ for 1D and 2D $\phi^4$-models, and for lattice sizes: $N=10\times 10$ (circles), $N=20\times 20$ (squares), $N=30\times 30$ (triangles) in the 2D case, and $N=900$ (rhombs) in the 1D case. The vertical dashed line indicates the phase transition point at $\bar{v}\simeq 2.2$.}
\label{var-chi}
\end{figure}

\begin{figure}[h]
 \centering
 \includegraphics[scale=0.45,keepaspectratio=true]{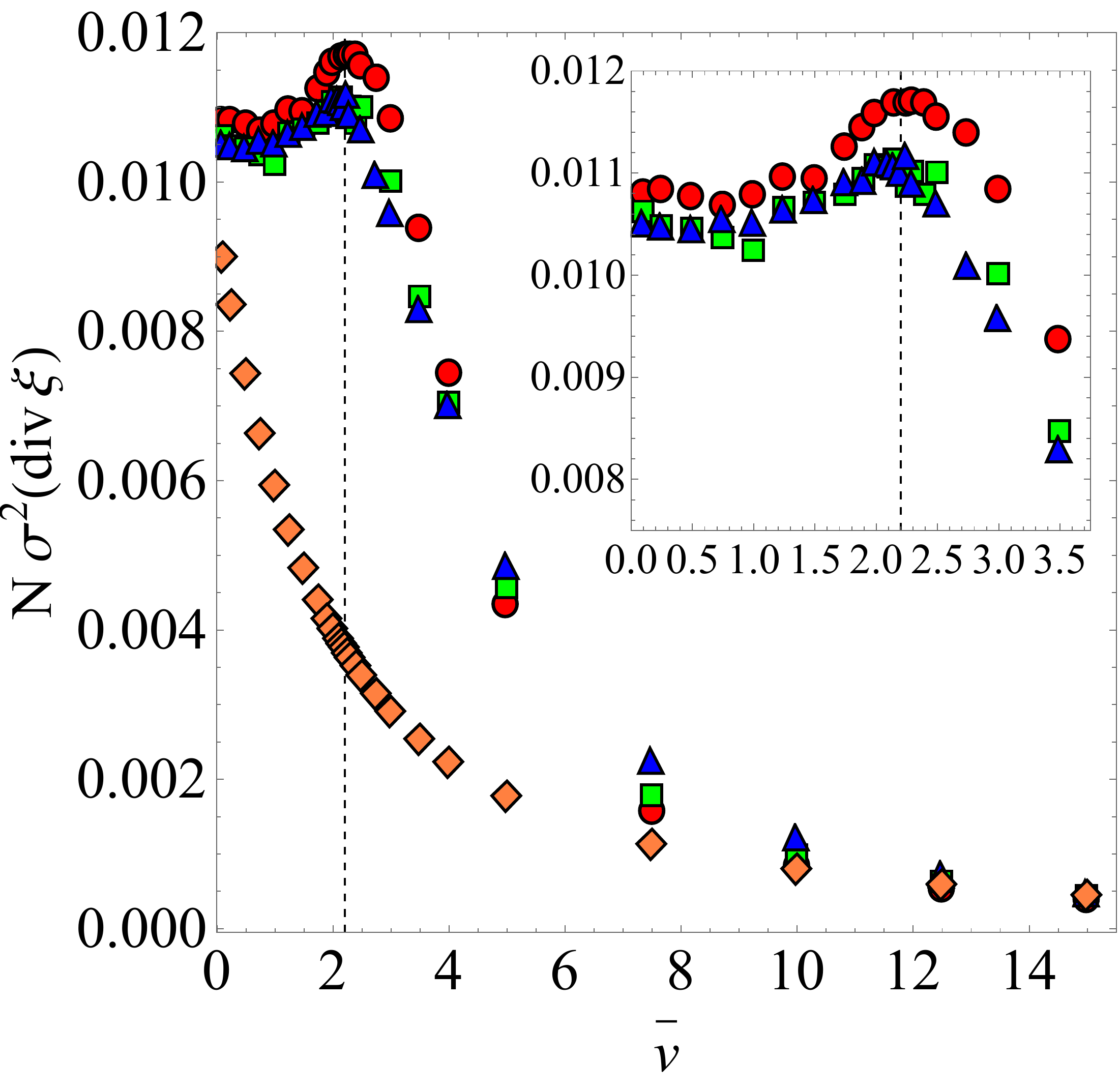}
\caption{(Color online) Variance of $\mathrm{div}\boldsymbol{\xi}$ vs. potential energy per degree of freedom $\bar{v}$ for 1D and 2D $\phi^4$-models, and for lattice sizes: $N=10\times 10$ (circles), $N=20\times 20$ (squares), $N=30\times 30$ (triangles) in the 2D case, and $N=900$ (rhombs) in the 1D case. The vertical dashed line indicates the phase transition point at $\bar{v}\simeq 2.2$.}
\label{var-diverg}
\end{figure}

The first variation formula  for the induced measure of the Riemannian area  $\mathrm{d}\sigma$ along the flow $q(v)$ reads \cite{LeeGeometry}:
\begin{equation}
\mathrm{d}\sigma(q(v+\epsilon_v))=\left(1-\epsilon_v\chi M_1\right)\mathrm{d}\sigma(q(v))+o(\epsilon_v)
\end{equation}
where $M_1$ is $N$ times the mean curvature of $\Sigma_v^{V_N}$ and is given by
\begin{equation}
M_1=-\mathrm{div}\left(\dfrac{\nabla V_{N}}{\|\nabla V_{N}\|}\right).
\end{equation}
In analogy with the case of phase space, applying the Leibnitz rule, the first variation formula for the measure of the microcanonical configurational area is
\begin{equation}\label{muarea}
\begin{split}
&\mathrm{d}\mu(q(v+\epsilon_v))=\chi(q(v+\epsilon_v))\mathrm{d}\sigma(v+\epsilon_v)=\\
&=\left[1+\epsilon_v\left(-\chi M_1+\dfrac{(\nabla^i V)}{\|\nabla V\|}\nabla_i\chi\right)\right]\mathrm{d}\mu=\\
&=\left(1+\epsilon_v\mathrm{div}\boldsymbol{\xi}\right)\mathrm{d}\mu(q(v))\ .
\end{split}
\end{equation}
Then,  the variances $\sigma^2(\chi)=\langle\chi^2\rangle_{\Sigma_v^{V_N}} - \langle\chi\rangle^2_{\Sigma_v^{V_N}}$ and $\sigma^2(\mathrm{div}\boldsymbol{\xi})=\langle(\mathrm{div}\boldsymbol{\xi})^2\rangle_{\Sigma_v^{V_N}} - \langle(\mathrm{div}\boldsymbol{\xi})\rangle^2_{\Sigma_v}$ have been numerically computed along the mentioned  Monte Carlo Markov Chain. These are functions of $N$ and of the specific potential energy ${\overline v}=V_N/N$.
The outcomes, reported in Figs. \ref{var-chi} and \ref{var-diverg}, also for configuration space show very different patterns in the $1D$ and $2D$ cases: monotonic  for the $1D$ case, non-monotonic displaying  cuspy points at ${\overline v}={\overline v}_c$ (the phase transition point) of $\sigma^2(\chi)$ and of $\sigma^2(\mathrm{div}\boldsymbol{\xi})$ for  the $2D$ case. 
Now, the variance of $\chi =1/\Vert\nabla V_N\Vert$ is a measure of the total dishomogeneity of the distance between nearby potential level sets. Moreover, the configurational volume (last integral in the r.h.s. of Eq.\eqref{mu-ent}) 
\begin{equation}
\Omega(v) = \int_{\Sigma_v^{V_N} }\chi\,\,\D\sigma_g\ ,
\end{equation}
likewise to what has been discussed for  phase space, under a rescaling of the metric $g$ of configuration space becomes
\begin{equation}
\Omega(v)=\int_{\Sigma_v^{V_N}}\,\D\sigma_{\widetilde{g}}
\end{equation}
so that the divergence of the vector field $\boldsymbol{\xi}$ in the configuration space endowed with the metric $g$ can be interpreted
as the opposite of $\widetilde{M}_{1}$ which $N$ times the mean curvature of the potential level sets $\Sigma_v^{V_N}$ embedded in the configuration space endowed with the metric $\widetilde{g}$, that is,
$\mathrm{div}_g\boldsymbol{\xi} = - \widetilde{M}_1$. Hence $\sigma^2(\mathrm{div}\boldsymbol{\xi})$ is the same as the variance of the mean curvature of the $\Sigma_v$ endowed with the metric $\widetilde{g}$.  By the same token discussed for the phase space, the cuspy patterns of $\sigma^2(\chi)$ and of $\sigma^2(\mathrm{div}\boldsymbol{\xi})$ can be ascribed to the formation of a ``neck" on the $\{\Sigma_v^{V_N}\}_{v\in\mathbb{R}}$ for  ${\overline v}<{\overline v}_c$. This neck appears as the restriction to configuration space of the neck on the energy level sets that exist below the phase transition point, as discussed in Section \ref{dynH}. At the same time, because of the above recalled geometrical and topological triviality related to the kinetic energy part of the Hamiltonian function, the necks of the potential level sets are at the grounds of the necks of the energy level sets.

\section{Discussion }\label{conclus}  
In spite of the absence of critical points of the potential $V_N(q)$ of the $\phi^4$ model [Eq.\eqref{potfi4}], also the phase transition occurring in this model stems from a topological change of configuration space submanifolds.  In particular, we have here shown that this transition stems from an \textit{asymptotic} change of topology, of both the $\Sigma_E^{{\cal H}_{N}}$ and $\Sigma_v^{V_N}$, in correspondence with the transition potential energy density ${\overline v}_c$. This paves the way to a more general formulation of the topological theory of phase transitions once a basic assumption of the theory is made explicit also in the $N\to\infty$ limit. Accordingly, in the theorems of Refs. \cite{prl1,TH1,TH2} the assumption of asymptotic diffeomorphicity of the $\{\Sigma_v^{V_n} \}_{n\in\mathbb{N}}$ has to be added to the hypothesis of finite-$N$-diffeomorphicity. Loosely speaking, while in a topological setting, this situation is reminiscent of the Yang-Lee theory, where only asymptotically in $N$ there exists the possibility of nonuniform convergence of the sequences of thermodynamic functions, and thus of the appearance of their nonanalytic behavior.

\subsection{Asymptotic Diffeomorphicity} 
The content of the present section is aimed at pointing out that ``asymptotic diffeomorphism'' is not a mathematically empty wording and that a proper definition can be naturally provided.
 This can be done by observing that a vector valued function of several variables, $f:\mathbb{R}^n\rightarrow\mathbb{R}^n$, is of differentiability class ${\cal{C}}^{l}$ if all the partial derivatives $(\partial^lf/\partial x_{i_1}^{l_1}\dots\partial x_{i_k}^{l_k})$ exist and are continuous, where each of $i_1,\dots,i_k$ is an integer between $1$ and $n$ and each $l_1,\dots,l_k$ is an integer between $0$ and $l$, and $l_1+\dots +l_k=l$.
Then, by taking advantage of a known analytic representation of the diffeomorphism $\boldsymbol\xi_N:\Sigma_{v}^{V_N}\subset\mathbb{R}^{N}\rightarrow\Sigma_{v^\prime}^{V_N}\subset\mathbb{R}^{N}$ (see below), and by introducing a suitable norm that contains all the derivatives up to $(\partial^l\boldsymbol\xi_N/\partial x_{i_1}^{l_1}\dots\partial x_{i_k}^{l_k})$, uniform convergence in $N$ of the sequence $\{\boldsymbol\xi_N\}_{N\in\mathbb{N}}$ - and thus asymptotic diffeomorphicity in some class ${\cal{C}}^{l}$ - can be naturally defined. 

At any fixed $N\in \mathbb{N}$, for confining potentials and in absence of critical points $V_N$, the level sets $\Sigma_v^{V_N}$ are non singular $(N-1)$-dimensional hypersurfaces in $\mathbb{R}^N$.

Consider now an open set of $v$-values $I\subseteq \mathbb{R}$ such that the cylindrical subset of configuration space
\begin{equation}
\Gamma_I^N=\bigcup_{{v}\in {I}}\Sigma^{V_N}_{v}
\end{equation}
contains only non-singular level sets, that is,  $V$ has no critical points for  any ${v}\in {I}$.

Then for any interval $\left[{v}_0,{v}_1\right]={I}_0\subset{I}$ any two level sets in $\Gamma^N_{{I}_0}$ are diffeomorphic 
(see \cite{hirsch}) under the action of an explicitly known diffeomorphism, as already shown in Eq.(\ref{hirschflow}).
This is given by the integral lines of the vector field which now reads $\boldsymbol{\xi}_N={\nabla{V}_N}/{\|\nabla{V}_N\|^2}$, that is
\begin{equation}
\frac{d{\bf q}}{dv} =\dfrac{\nabla{V}_N}{\|\nabla{V}_N\|^2} \qquad
\end{equation} 
with any initial condition $\mathbf{q}_0\in\Sigma_{{v}_0}^n$.

In order to characterize the asymptotic breaking of diffeomorphicity we introduce a norm for the
$\boldsymbol{\xi}_N$ that allows to compare the diffeomorphisms at different dimensions
\begin{equation}
\|\boldsymbol{\xi}_N \|_{C^k(\Gamma^N_{{I}_0})}=\sup_{\mathbf{q}_0\in\Gamma^N_{{I}_0}}\|\boldsymbol{\xi}_N\|+
\dfrac{1}{N}\sum_{l=1}^{k}\sum_{\{ i_k\}}\sum_{j=1}^N\|{\nabla^l_{\{ i_k\}}}{\xi}_j\|_{\Gamma^N_{{I}_0}}
\end{equation}
where $\{ i_k\}$ is a multi-index and $\|{\nabla^l_{\{ i_k\}}}{\xi}_j\|_{\Gamma^N_{{I}_0}}$ is the norm of the $l$-th differential operator 
\begin{equation}
\|{\nabla^l_{\{ i_k\}}}{\xi}_j\|_{\Gamma^N_{{I}_0}}=\sup_{\mathbf{q}_0\in\Gamma^N_{{I}_0}}\left\vert \dfrac{\partial^l{\xi}_j}{\partial q_{i_1}^{l_1}\dots\partial q_{i_k}^{l_k}}\right\vert \ 
\label{norma}
\end{equation}
with $l_1+\dots +l_k=l$.

We say that the sequence of families of manifolds $\left\{\Gamma^N_{{I}_0}\right\}_{N\in\mathbb{N}}$ 
asymptotically preserves the $C^{k}$-diffeomorphicity  among the hypersurfaces of each family if there exists $B\in\mathbb{R}^{+}$ such
that
\begin{equation}
\label{eq:AsympDiffCk}
\|\boldsymbol{\xi}_N \|_{C^k\left(\Gamma^N_{{I}_0}\right)}\leq B<+\infty \qquad \forall N\in \mathbb{N}.
\end{equation}
This condition implies  
$\|\nabla{V}_N\|=\|{\boldsymbol\xi}_N\|^{-1}\geq 1/B=C>0$ for each $\mathbf{q}_0\in\Gamma_{{I}_0}^N$ and all $N\in\mathbb{N}$, thus  excluding the existence of asymptotic critical points (that is $\|\nabla{V}_N\|\rightarrow 0$ for $N\rightarrow\infty$).

Moreover, using $\sum_i \Vert X_i\Vert \ge \Vert \sum_i X_i\Vert$, from Eq. \eqref{norma} we can write at the lowest order

\begin{equation}
\sum_{i,j=1}^N\Vert \partial_i\xi_j\Vert \ge \left\Vert \sum_{i,j=1}^N \partial_i \dfrac{\partial_j V_N}{\Vert\nabla V_N\Vert^2} \right\Vert
\end{equation}
where $\partial_i=\partial/\partial q^i$.
Then at any given point $\mathbf{q}_0\in\Gamma^N_{{I}_0}$ we build the quadratic form
\begin{equation}
 \left\Vert \sum_{i,j=1}^N \left(\partial_i \dfrac{\partial_j V_N}{\Vert\nabla V_N\Vert^2}\right) u_i u_j \right\Vert
\end{equation}
by using a normalised vector ${\bf u}$ tangent at $\mathbf{q}_0$ to a $\Sigma_v^n\subset \Gamma^N_{{I}_0}$. With implicit summation on repeated indices we get
\begin{eqnarray}
 &&\left\Vert\left( \partial_i \dfrac{\partial_j V_N}{\Vert\nabla V_N\Vert^2}\right) u_i u_j \right\Vert  \label{Qform}\\
 &=& \left\Vert \dfrac{1}{\Vert\nabla V_N\Vert}\left(\partial_i \dfrac{\partial_j V_N}{\Vert\nabla V_N\Vert}\right) u_i u_j 
 +  \dfrac{\partial_j V_N}{\Vert\nabla V_N\Vert}\partial_i \left( \dfrac{1}{\Vert\nabla V_N\Vert}\right) u_i u_j  \right\Vert \nonumber \\
 &=& \left\Vert \dfrac{1}{\Vert\nabla V_N\Vert}\left(\partial_i \dfrac{\partial_j V_N}{\Vert\nabla V_N\Vert}\right) u_i u_j \right\Vert  \nonumber
\end{eqnarray}
where we have used the orthogonality, at any given point $\mathbf{q}_0$, between the vectors ${\bf u}$ and  ${\cal N}=( \partial_1 V_N/\Vert\nabla V_N \Vert,\dots,\partial_N V_N/\Vert\nabla V_N\Vert$ which are tangent and normal to $\Sigma_v^{V_N}$, respectively.

If we now consider the Weingarten map (shape operator) of $\Sigma_v^{V_N}$ \cite{thorpe} at $\mathbf{q}_0$
\begin{equation}
L_{\mathbf{q}_0}({\bf u}) = - L_{{\bf u}}{\cal N} = - (\nabla {\cal N}_1\cdot {\bf u},\dots,\nabla {\cal N}_N\cdot {\bf u})
\label{shape}
\end{equation}
we see that the quadratic form $k({\bf u}) = \langle {\bf u}, L({\bf u})\rangle$ coincides with that one built in Eq.\eqref{Qform} (last term). The quantity $k({\bf u})$ is the normal curvature of the level set $\Sigma_v^n$ at any given point. Thus asymptotic diffeomorphicity, defined as uniform boundedness of the above given norm, entails the uniform boundedness of $k({\bf u})$ and,
consequently, of all the principal curvatures of $\Sigma_v^{V_N}$. In other words, this rules out the sequences of manifolds $\{{\cal{M}}_{N}\}_{N\in\mathbb{N}}$ and $\{{\cal{S}}_{N}\}_{N\in\mathbb{N}}$ mentioned in Section \ref{tophyp}.
This is particularly evident in the G-H limit of \autoref{GHlimt} where the transverse radius of curvature of a thinner and thinner neck vanishes asymptotically, thus making the corresponding principal curvature divergent. This is to illustrate that the above given definition of asymptotic diffeomorphicity is sound and consistent.

\subsection{Conclusion}
In conclusion, let us remark again that under the assumption of diffeomorphicity at any arbitrary \textit{finite} $N\in\mathbb{N}$ of any pair of $\Sigma_{v}^{V_N}\subset {\mathbb R}^N$ - which is not equivalent to the absence of critical points, as discussed at the beginning of Section II - the two basic theorems in \cite{prl1,TH1,TH2} derived the uniform convergence of Helmholtz free energy at least in the ${\cal C}^2({\mathbb R})$ differentiability class. Consequently, the occurrence of a phase transition would necessarily require the loss of diffeomorphicity  of the level sets $\Sigma_{v}^{V_N}$. This is falsified by the $2D$ lattice $\phi^4$-model, because its phase transition takes place without a loss of diffeomorphicity of the $\Sigma_{v}^n$, and this is motivated by the absence of critical points of the potential function. 
However, the present work indicates what is missing in the hypotheses of the theorems above. In fact,  we have seen that the phase transition of the $2D$ lattice $\phi^4$-model corresponds to an asymptotic ($N\rightarrow\infty$)  breaking of the topological transitivity of phase space and of configuration space level sets, $\Sigma_{E}^{{\cal H}_{N}}$ and $\Sigma_{v}^{V_N}$ respectively. Thus the way to fix the problem appears to extend the basic hypothesis of the theorems in \cite{prl1,TH1,TH2} by encompassing also asymptotic diffeomorphicity of the $\{\Sigma_{v}^{V_N}\}$, because in this way the $\phi^4$-model will no longer fulfil the hypotheses of the theorems, and thus will no longer be a counterexample of the theory. And this is appropriate because the phase transition of the $\phi^4$-model actually corresponds to a major topological change of submanifolds of both phase space and configuration space.

In this context it is worth mentioning that with a completely different approach also the phase transition of the $2D$ Ising model (which is of the same universality class of the $2D$ lattice $\phi^4$ model) is found to correspond to an asymptotic change of topology of suitable manifolds. This is found by proving that the analytic index of a given elliptic operator - acting among smooth sections of a vector bundle defined on a state manifold -  makes an integer jump at the transition temperature of the $2D$ Ising model \cite{rasetti1,rasetti2}. Hence the asymptotic change of topology of sections of the mentioned vector bundle stems from the Atiyah-Singer index theorem which states that the analytic index is equal to a topological index \cite{nakahara}.

\vfill
\section{Appendix}
A link among the topology of (specific) energy level sets and the topology of configuration
space accessible to the system can be established, and this is possible as the Topological Theory of phase transitions is (in its present formulation) restricted to systems whose microscopic dynamics is described by Hamiltonian of the form $\mathcal{H}_N(p,q)=\sum_{i=1}^N p_i^2/2+V_N(q_1,...,q_N)$ with short-range potentials bounded from below (we can suppose that $\bv_{min}=0$ for all $N$).\\
Hence, (using for the moment a cumbersome notation for the sake of clarity) the level sets $\LevelSetfunc{E}{{\mathcal{H}}_N}$ of the energy function ${\mathcal{H}}_N$  can be given by the disjoint union of a trivial unitary sphere bundle (representing the phase space region where the kinetic energy does not vanish) and the hypersurface in configuration space where the potential energy takes total energy value.
In fact we can define a map between points  $\left(p_1,...,p_N,q_1,...,q_N\right)\in\Sigma^{\mathcal{H}_N}_{E}$ and points in $M_{E}^{V_N}\times\mathbb{S}^{N-1}\bigsqcup\LevelSetfunc{{E}}{{V}_N}$
\begin{equation}
\begin{cases}
& q_{i}=x_{i}\qquad \forall i=1,...,N\\
& p_{1}=\left[2 \left(E-V(q_1,...q_N)\right)\right]^{1/2}\cos(\theta_1)\\
& p_{2}=\left[2 \left(E-V(q_1,...q_N)\right)\right]^{1/2}\sin(\theta_1)\cos(\theta_2)\\
& .............................................\\
&p_{N-1}=\left[2 \left(E-V(q_1,...q_N)\right)\right]^{1/2}\sin(\theta_1)\sin(\theta_2)\cdots\times\\
&\times\sin(\theta_{N-2})\cos(\theta_{N-1})\\
&p_{N}=\left[2 \left(E-V(q_1,...q_N)\right)\right]^{1/2}\sin(\theta_1)\sin(\theta_2)\cdots\times\\
&\times\sin(\theta_{N-2})\sin(\theta_{N-1})\\
\end{cases}
\end{equation}
where the points $(x_{1},...,x_{N})\in M_{E}^{V_N}\bigsqcup\LevelSetfunc{{E}}{{V}_N}$ and the angles $\theta_{1},....\theta_{N-2}\in[0,\pi)$ and $\theta_{N-1}\in[0,2\pi]$ are a parametrization of the unitary $(N-1)$-sphere $\mathbb{S}^{N-1}$.
From this it follows that
\begin{equation}
\LevelSetfunc{E}{{\mathcal{H}}_N} {\mathrm{\ homeomorphic\  to}\ } \Mfunc{E}{{V}_N}\times \mathbb{S}^{N-1}\,\,\,\bigsqcup\,\,\,\LevelSetfunc{{E}}{{V}_N}
\label{eq:Structure_sigmaE}
\end{equation}
where $\mathbb{S}^n$ is the $n$-dimensional unitary sphere and 
\begin{equation}
\begin{split}
&\Mfunc{c}{f}=\left\{x\in\mathrm{Dom}(f)|f(x)< c\right\}, \\ \\
 &\LevelSetfunc{c}{f}=\left\{x\in\mathrm{Dom}(f)|f(x)=c\right\}.
 \end{split}
\end{equation}
The idea that finite $N$ topology, and "asymptotic topology" as well, of $\LevelSetfunc{{E}}{{\mathcal{H}}_N}$ is affected by the  topology of the accessible region of configuration space is suggested by the \textit{K\"unneth formula}: if $H_{k}(X)$ is the $k$-th homological group of the topological space $X$ on the  field $\mathbb{F}$ then
\begin{equation}
H_{k}(X\times Y;\mathbb{F})\simeq\bigoplus_{i+j=k} H_{i}(X;\mathbb{F})\,\,\otimes\,\, H_{j}(Y;\mathbb{F})\,\,.
\end{equation}
Moreover, as $H_{k}\left(\sqcup_{i=1}^N X_{i},\mathbb{F}\right)=\bigoplus_{i}^{N}H_{k}(X_{i},\mathbb{F})$, it follows that:
\begin{equation}
\begin{split}
&H_{k}\left(\LevelSetfunc{{E}}{{\mathcal{H}}_N},\mathbb{R}\right) \label{homolog}\\
&\simeq\bigoplus_{i+j=k} H_{i}\left(\Mfunc{{E}}{{V}_N};\mathbb{R}\right)\otimes\,\, H_{j}\left(\mathbb{S}^{N-1};\mathbb{R}\right)\oplus H_{k}\left(\LevelSetfunc{{E}}{{V}_N};\mathbb{R}\right)\\
&\simeq  H_{k-(N-1)}\left(\Mfunc{{E}}{{V}_N};\mathbb{R}\right)\otimes \mathbb{R}\oplus H_{k}\left(\Mfunc{{E}}{{V}_N};\mathbb{R}\right)\otimes \mathbb{R}\\
&\oplus H_{k}\left(\LevelSetfunc{{E}}{{V}_N};\mathbb{R}\right) 
\end{split}
\end{equation}
the r.h.s. of Eq.\eqref{homolog} shows that the topological changes of $\LevelSetfunc{{E}}{{\mathcal{H}}_N}$ only stem from the topological changes in configuration space.



\begin{thebibliography}{99}
 %
\bibitem{PettiniBook} M. Pettini, {\it Geometry and Topology in Hamiltonian
		Dynamics and Statistical Mechanics}, IAM Series n. 33,
		(Springer-Verlag New York, 2007).

\bibitem{CCCP} L. Caiani, L. Casetti, C. Clementi, and M. Pettini, {\it Geometry of Dynamics, Lyapunov Exponents, and Phase Transitions}, Phys. Rev. Lett. \textbf{79},  4361 (1997).

\bibitem{CSCFP} M. Cerruti-Sola, G. Ciraolo, R. Franzosi and M. Pettini, {\it Riemannian geometry of
Hamiltonian chaos: Hints for a general theory}, Phys. Rev. E \textbf{78}, 046205 (2008).

\bibitem{CCCPPG} L. Caiani and L. Casetti and C. Clementi and G. Pettini and M. Pettini and R. Gatto, {\it  Geometry of dynamics and phase transitions in classical lattice $\phi^4$ theories}, Phys.Rev. E\textbf{57}, 3886 (1998).

\bibitem{physrep}  L. Casetti, M. Pettini, and E.G.D. Cohen, {\it  Geometric approach to Hamiltonian dynamics
and statistical mechanics}, Physics Reports \textbf{337}, 237 (2000).

\bibitem{prl1} R. Franzosi and M. Pettini, {\it Theorem on the Origin of Phase Transitions}, Phys. Rev. Lett. \textbf{92}, 060601 (2004).

\bibitem{TH1} R. Franzosi, L. Spinelli and M. Pettini, {\it Topology and phase transitions I. Preliminary results}, Nuclear Physics B \textbf{782}, 189 (2007).

 \bibitem{TH2} R. Franzosi and M. Pettini, {\it Topology and phase transitions II. Theorem on a necessary relation}, Nuclear Physics B \textbf{782}, 219 (2007).

\bibitem{FF} R. Franzosi, \textit{Microcanonical entropy and dynamical measure of temperature for systems
with two first integrals}, J. Stat. Phys. \textbf{143}, 824 (2011); R. Franzosi, \textit{Geometric microcanonical
thermodynamics for systems with first integrals}, Phys. Rev. E\textbf{85}, R050101 (2012).

\bibitem{carlsson1} G. Carlsson, J. Gorham, M. Kahle, and J. Mason, {\it Computational topology for configuration spaces of hard disks}, Phys. Rev. E\textbf{85},  011303 (2012).

\bibitem{barish} Y. Baryshnikov, P. Bubenik, and M. Kahle, {\it Min-type Morse theory
for configuration spaces of hard spheres}, Int. Math. Res. Notices, doi:10.1093/imrn/rnt012, (2013).

\bibitem{brody} D.C. Brody, D.W. Hook, and L. P. Hughston, {\it Quantum phase transitions
without thermodynamic limits}, Proc. Roy. Soc. A (London) \textbf{463},  2021 (2007).

\bibitem{BFS} P. Buonsante, R. Franzosi, A. Smerzi, {\it Phase transitions at high energy vindicate negative microcanonical
temperature}, arXiv:1506.01933.

\bibitem{volovik} G.E. Volovik, {\it Quantum phase transitions from topology in momentum space}, 
in {\sl Quantum analogues : from phase transitions to black holes and cosmology}, p.31-73 (Springer, Berlin Heidelberg, 2007).

\bibitem{angelani} L. Angelani, R. Di Leonardo, G. Parisi, and G. Ruocco, \textit{Topological description of the aging dynamics in simple glasses}, Phys. Rev. Lett. \textbf{87}, 055502 (2001).

\bibitem{stillinger} P.G. Debenedetti, and F. H. Stillinger, \textit{Supercooled liquids and the glass
transition}, Nature \textbf{410}, 259 (2001) .

\bibitem{risau} S. Risau-Gusman, A.C. Ribeiro-Teixeira, and D.A. Stariolo, \textit{Topology,
phase transitions, and the spherical model},  Phys. Rev. Lett. \textbf{95}, 145702 (2005).

\bibitem{schilling} D.A. Garanin, , R. Schilling, and A. Scala, \textit{Saddle index properties, singular topology,
and its relation to thermodynamic singularities for a $\phi^4$ mean-field model}, Phys. Rev. E\textbf{70}, 036125 (2004). 

\bibitem{fernando} F.A.N. Santos, L.C.B. da Silva, and M.D. Coutinho-Filho, {\it Topological approach to microcanonical thermodynamics and phase transition of interacting classical spins}, Journal of Statistical Mechanics: Theory and Experiment  {\bf 2017}, 013202 (2017).

\bibitem{grinza} P. Grinza, and A. Mossa, \textit{Topological origin of the phase transition in a
model of DNA denaturation}, Phys. Rev. Lett. \textbf{92},  158102 (2004).

\bibitem{becker} O.M. Becker,  and M. Karplus, \textit{The topology of multidimensional potential energy surfaces: theory and application to peptide structure and kinetics},  J. Chem. Phys. \textbf{106}, 1495 (1997).

\bibitem{bachmann} M. Bachmann, {\it Thermodynamics and Statistical Mechanics of Macromolecular Systems}, (Cambridge University Press, New York, 2014).

\bibitem{Carlsson} G. Carlsson and A. Zomorodian, {\it Persistent homology - a survey}, Discrete Comput. Geom. \textbf{33}, 249 (2005); G. Carlsson, {\it Topology and data}, Bull. Am. Math. Soc. \textbf{2}, 255 (2009).

\bibitem{noi} I. Donato, M. Gori, M. Pettini, G. Petri, S. De Nigris, R. Franzosi, and F. Vaccarino
\textit{Persistent Homology analysis of Phase Transitions}, Phys. Rev. E\textbf{93}, 052138 (2016);
and references therein for persistent homology.

\bibitem{kastner} M. Kastner and D. Mehta, {\it Phase Transitions Detached from Stationary Points of the Energy Landscape}, Phys. Rev. Lett. \textbf{107}, 160602 (2011).

\bibitem{DLR} D. Ruelle,  {\it Thermodynamic Formalism}, Encyclopaedia of Mathematics and Its Applications, (Addison?Wesley, New York, 1978); 

\bibitem{georgii} H.O. Georgii,  {\it Gibbs Measures and Phase Transitions}, (Walter de Gruyter, Berlin, 1988).

\bibitem{sormani} Christina Sormani, {\it How Riemannian Manifolds Converge}, In: Dai X., Rong X. (eds) Metric and Differential Geometry. Progress in Mathematics, vol 297. Birkh\"auser, Basel (2012).
 
\bibitem{CCP} L. Caiani, L. Casetti, and M. Pettini, {\it Hamiltonian dynamics of the two-dimensional lattice $\phi^4$ model}, J.Phys.A: Math.Gen. \textbf{31}, 3357 (1998).

\bibitem{Lapo} L. Casetti, {\it Efficient symplectic algorithms for numerical simulations of Hamiltonian flows},
Physica Scripta \textbf{51}, 29 (1995).

\bibitem{nota} The parameters chosen here are the same of Ref.\protect\cite{CCP} where the critical energy density is shifted by a constant vaue of $10$.

\bibitem{nota3} The $N\to\infty$ extrapolation is safe because increasing $N$ essentially amounts to gluing togheter identical replicas of smaller lattices.

\bibitem{kac} M. Kac, {\it On the notion of recurrence in discrete stochastic processes}, Bull. Amer. Math. Soc. {\bf 53}, 1002 - 1010 (1947).

\bibitem{notaergodicity} After the Poincar\'e-Fermi theorem, a non-integrable Hamiltonian system with $N\ge 3$ is {\it bona-fide} ergodic. Especially at large $N$, no hindrance to ergodicity can be given by KAM theorem (see \protect{\cite{PettiniBook} Chapter 2}).

\bibitem{varandas} P. Varandas, {\it A version of Kac's lemma on first return times for suspension flows}, Stoch. Dyn. {\bf 16}, 1660002 (2016).

\bibitem{moser} J.K. Moser, {\it Lectures on Hamiltonian systems}, Mem. Am. Math. Soc. {\bf 81}, 1-60 (1968).

\bibitem{benettin} G. Benettin, and A. Giorgilli, {\it On the Hamiltonian interpolation of near-to-the identity symplectic mappings with application to symplectic integration algorithms}, J. Stat. Phys. {\bf 73}, 1117 - 1143 (1994).

\bibitem{toptran} For a general introduction to topological transitivity of flows and related properties, see:
J. M. Alongi and G. S. Nelson, {\it Recurrence and Topology}, Graduate Studies in Mathematics, Volume 85, (American Mathematical Society, Providence, Rhode Island, 2007). 

\bibitem{ledoux} M. Ledoux, {\it The Concentration of Measure Phenomenon}, Mathematical Surveys and Monographs, Vol. 89, (American Mathematical Society, Providence, Rhode Island, 2001). 

\bibitem{hirsch} M.W. Hirsch, {\it Differential Topology}, (Springer, New York 1976).

\bibitem{gromov} M. Gromov,  {\it Isoperimetric inequalities in Riemannian manifolds}, Appendix I to V. D. Milman and G. Schechtman,  {\it Asymptotic Theory of Finite Dimensional Normed Spaces}, Lecture Notes Math., vol. 1200, Springer-Verlag (1986).

\bibitem{bayle} V. Bayle,  {\it Propri\'et\'es de concavit\'e du profil isop\'erim\'etrique et applications}, Diss. Universit? Joseph-Fourier-Grenoble I, (2003).

\bibitem{morgan} F. Morgan,  {\it Manifolds with density}, Notices of the AMS, 853-858 (2005).

\bibitem{LeeGeometry} Lee, J.M., {\it Manifolds and Differential Geometry}, (American Mathematical Society, 2009).

\bibitem{thorpe} J.A. Thorpe, {\it Elementary Topics in Differential Geometry}, (Springer-Verlag, New York 1979), p. 55.


\bibitem{rasetti1} M.Rasetti, {\it Topological Concepts in Phase Transition Theory}, in
{\it Differential Geometric Methods in Mathematical Physics}, H. D. D\"obner (ed.)
(Springer-Verlag, New York, 1979).

\bibitem{rasetti2} M.Rasetti, {\it Structural Stability in Statistical Mechanics}, p. 159, W. G\"uttinger et al. (eds.), {\it Structural Stability in Physics}, (Springer-Verlag, Berlin, Heidelberg 1979).

\bibitem{nakahara} M. Nakahara, {\it Geometry, Topology and Physics}, (Adam Hilger, Bristol, 1991).

\end{thebibliography}
\end{document}